\documentclass{journal_stylefiles/ar-1col-S2O}

\usepackage{fix-cm}
\usepackage[numbers]{natbib}
\usepackage{url}
\usepackage{amsmath}
\usepackage{ifoddpage}

\usepackage{tabularx}
\usepackage{array}
\usepackage{booktabs}
\usepackage{makecell}
\usepackage{amsthm}
\usepackage{mathtools}

\usepackage[colorlinks=false, citecolor=black, linkcolor=black, urlcolor=black]{hyperref}

\usepackage{amsfonts}
\usepackage{amssymb}
\usepackage{comment}
\usepackage{enumitem}
\usepackage{framed}

\newcolumntype{Y}{>{\centering\arraybackslash}X}
\newcommand{\tableshift}{%
  \checkoddpage
  \ifoddpage
    \hspace*{-1.0cm}
  \else
    \hspace*{-4.0cm}
  \fi
}

\newif\ifshowcomments
\showcommentstrue
\ifshowcomments
  \newcommand{\matt}[1]{\textcolor{blue}{[MS: #1]}}
  \newcommand{\martin}[1]{\textcolor{red}{[MM: #1]}}
  \newcommand{\todo}[1]{\textcolor{red}{[TODO: #1]}}
\else
  \newcommand{\matt}[1]{}
  \newcommand{\martin}[1]{}
  \newcommand{\todo}[1]{}
\fi

\jname{Accepted for publication in
Annual Review of Control, Robotics, and Autonomous Systems.}
\jvol{10}
\jyear{xxxx}
\doi{10.1146/annurev-control-050825-104421 
\newline \newline
\emph{Accepted author manuscript, prior to copyediting, illustration, and typesetting by Annual Reviews}
}

\begin{document}

\markboth{Smart et al.}{Dynamical principles of habituation}

\title{Dynamical principles of habituation across substrates and scales}

\author{Matthew Smart,$^{1,2}$ Stanislav Y. Shvartsman,$^{1,2,3}$ and Martin M\"onnigmann$^4$
\affil{$^1$Lewis-Sigler Institute for Integrative Genomics, Princeton University, Princeton, NJ, USA; mattsmart@princeton.edu}
\affil{$^2$Center for Computational Biology, Flatiron Institute, New York, NY, USA}
\affil{$^3$Department of Molecular Biology, Princeton University, Princeton, NJ, USA; stas@princeton.edu}
\affil{$^4$Department of Mechanical Engineering, Ruhr-Universit\"at Bochum, Bochum, Germany; martin.moennigmann@rub.de}}

\begin{abstract}
Habituation is a basic form of learning in which a system's response to repeated stimulation progressively diminishes but eventually recovers when the stimulus is withheld. Long studied in animals, it has increasingly been observed in unicellular organisms and non-living devices such as electronic circuits and neuromorphic materials, suggesting underlying dynamical principles that recur across domains. 
This review asks what those principles are: given qualitative constraints imposed by habituation on a system's 
response, what is the minimal dynamical structure that satisfies them? We formalize the classical hallmarks of habituation as behavioral constraints on input--output behavior, show that linear time-invariant systems are structurally incompatible with these constraints, and construct nonlinear motifs---linear fading-memory dynamics composed with static nonlinearities---that exhibit the hallmarks across diverse settings. We relate these motifs to models of specific biological systems and to physical and algorithmic realizations, from analog circuits to transient computation in machine learning.
\end{abstract}


\maketitle

\begin{figure}[t!]
\includegraphics[width=0.99\linewidth]{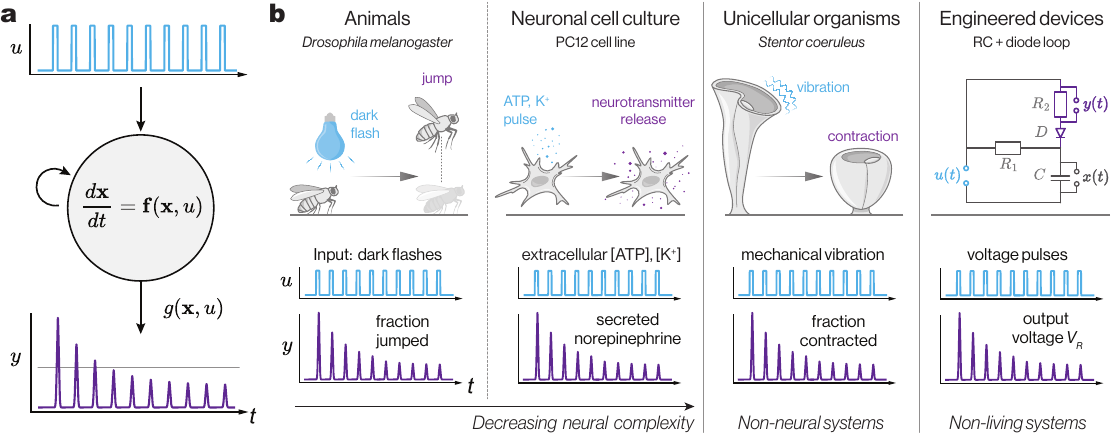}
\caption{
  \textbf{Habituation across domains and scales.}
  (a) A stimulus $u(t)$ drives internal dynamics $\dot{\mathbf{x}} = \mathbf{f}(\mathbf{x},u)$ whose output $y = g(\mathbf{x},u)$ attenuates across repeated presentations and recovers when stimulation is withheld.
  (b) Four concrete realizations spanning decreasing neural complexity, non-neural living systems, and non-living devices.
  \emph{Drosophila melanogaster}: dark flashes elicit a startle jump whose probability decreases over repeated trials.
  PC12 neuronal cell culture: extracellular ATP and K$^+$ pulses trigger norepinephrine secretion that diminishes with repetition.
  \emph{Stentor coeruleus}: a single-celled ciliate that contracts in response to mechanical stimuli. 
  RC circuit with diode loop: an engineered analog of a habituation motif, where the output voltage attenuates under repeated voltage pulses.
  Despite differences in mechanisms, all four systems exhibit the same qualitative attenuation-and-recovery dynamics.
  Adapted with permission (CC BY-NC-ND 4.0) from Smart et al.~\citep{smart2024pnas}. Circuit schematic in (b) redrawn after ref. \citep{Monnigmann2024CDC}.
}
\label{fig:fig1_overview}
\end{figure}

\section{INTRODUCTION}
\label{sec:intro}

Organisms are immersed in environments dense with input, most of it irrelevant. To act effectively under limited resources such as energy and attention, a system must filter salient signals from a benign, repetitive background. Habituation is perhaps the simplest solution to this problem and the most basic form of nonassociative learning: the response to a repeated, inconsequential stimulus progressively declines, but recovers once the stimulus is withheld \cite{Thompson2009}. 
Though habituation was first characterized in biology, the problem it solves (and, as we argue, the dynamical structure that solves it) is not specific to living systems. 
This review develops that claim from a systems and control perspective, asking what dynamical structure any habituating system must possess, regardless of domain.

While traditionally studied in neural contexts alongside other forms of learning, habituation does not require a nervous system (\textbf{Figure~\ref{fig:fig1_overview}}). Habituation-like behavioral plasticity was documented in unicellular organisms more than a century ago \cite{jennings1902studies, Jennings1906}, and is now widely recognized, to varying degrees, across the tree of life: not only in vertebrates \cite{davis1970effects, groves1970habituation, Randlett2019} and invertebrates \cite{Carew1972, kandel1976cellular, rankin1990caenorhabditis, Engel2009} but also in cultured neurons \cite{Mcfadden1990, Cheever1994}, in certain plants \cite{Gagliano2014} and fungi \cite{ortega1970phycomyces,applewhite1975learning}, and even in nonliving materials whose internal states are reshaped by repeated stimulation. Improved single-organism and single-device measurement has renewed interest in the aneural and nonliving settings, including in ciliates \cite{wood1969parametric, Eisenstein1982, Rajan2023}, slime molds \cite{Boisseau2016, Boussard2019}, and engineered devices \cite{Zhang2021resistance, wu2020_material_hab_volatile};
as the interpretation of what constitutes learning in single cells remains debated, we refer the reader to \cite{gershman2021reconsidering} for a careful appraisal, and \cite{gunawardena2022learning} for a broader perspective.
Across this breadth of domains --- spanning disparate molecular machinery, architectures, and scales --- habituation has long been described not by an equation but by a set of qualitative \emph{hallmarks}: verbal characteristics, such as the dependence of the response decrement on the frequency and intensity of stimulation, cataloged for animal behavior in the 1960s and refined more recently \cite{Thompson1966, rankin2009}. 
Only within the past few years have these hallmarks been given precise mathematical formulations, both as analytic behavioral criteria \cite{smart2024pnas, Monnigmann2024CDC} and as numerical filters for screening biochemical circuits \cite{eckert2024}.
That a phenomenon recurs so faithfully across systems with so little in common in physical form is the central clue this review pursues: its hallmarks appear to reflect dynamical principles that are independent of domain.
This perspective is cybernetic in spirit \cite{Wiener1948cybernetics, ashby1956introductionCybernetics},
focusing on dynamical features and constraints shared across realizations rather than the particular mechanisms that implement them.
Habituation's filtering role also makes its breakdown consequential:
impaired habituation is a recurring feature of neurodevelopmental disorders \cite{Blok2022, Fenckova2019}, where a failure to attenuate familiar input is thought to degrade downstream processing, a cognitive ``firewall'' \cite{poon2006nonassociative} interpretation we return to at the end.
(Habituation should also be distinguished from the related phenomenon of \emph{adaptation}; 
see Section \ref{sec:adapt-vs-hab}.)

This review approaches habituation as a problem in dynamical systems and control. 
The premise is simple: biological systems must implement dynamical systems whose responses obey qualitative behavioral requirements \emph{robustly} across variation in parameters, initial conditions, and the precise form of the input. Habituation is an unusually clean instance of this problem. 
Its hallmarks are not data to be fit but specifications to be met, and the question they pose is structural: what is the \emph{minimal} dynamical structure whose transient response satisfies them, not for one trajectory but for an entire class of admissible inputs? 
Framed this way, habituation sits slightly to the side of the classical concerns of control. Tasks like disturbance rejection and reference tracking and filter design canonically constrain \emph{asymptotic} behavior, i.e., what the output does as $t \to \infty$. Habituation instead constrains the \emph{transient}: the requirements are on the shape of the signals, i.e., the history-dependent attenuation of successive responses, and not on the limit alone. Methods that involve transient behavior, such as optimal control and system identification, exist, but they apply to particular instances of time series data, such as a particular collection of input-output data or a particular target output behavior, respectively. The problem at hand here, in contrast, is not characterized by particular transient behaviors, but by qualitative constraints such as the requirement of a diminishing response and recovery in the case of habituation.
We therefore frame habituation to the systems and control community in that spirit: not as a problem already solved by existing methods, but as a well-posed example of a distinct class to which such methods should apply or be extended toward.

The review proceeds as follows.
Section~\ref{sec:formalize} makes the task of designing a habituating system precise, casting the hallmarks as behavioral constraints on a system's input--output operator and establishing a first structural result: no linear time-invariant system can habituate, so nonlinearity is a necessity rather than a modeling choice. 
Section~\ref{sec:adapt-vs-hab} contrasts habituation with adaptation, the better-studied phenomenon of asymptotic disturbance rejection, and argues that the two, while superficially related, are logically independent, with neither implying the other. 
Section~\ref{sec:families} surveys the dynamical systems proposed as models of habituation, organized by mathematical structure rather than biological or technical origin, and constructs a minimal motif --- linear fading-memory dynamics composed with a static nonlinearity --- directly from the hallmarks interpreted as design specifications, with fading memory emerging as the unifying thread. 
Section~\ref{sec:realizations} turns to physical and algorithmic realizations, from analog circuits and neuromorphic materials to fading-memory models in machine learning.
Section~\ref{sec:open} collects open problems in the context of systems and control, and Section~\ref{sec:conclusions} concludes.

\section{FORMALIZING CONSTRAINTS ON TRANSIENT RESPONSES}
\label{sec:formalize}

Finding a mathematical description of a system's observed behavior is the task of \emph{system identification} \citep{Ljung1999-SysIdTheoryForTheUser, Ljung2008-PerspectivesOnSystemsIdentification}. In its usual engineering form, the problem is quantitative: given measured input--output time series from a particular physical system, find a model whose trajectories reproduce them, typically by selecting one or a few competing model structures and tuning their parameters to optimize a goodness-of-fit. 
The problem addressed here differs in two fundamental ways:

\begin{quote}
\emph{1. The desired behavior is not described by time-series of a particular instance of a system, but by qualitative rules.} 

\vspace{0.2cm}
\emph{2. The goal is not to find a particular model for specific data but a minimal model that collects the core features that dynamical systems capable of implementing the rules have in common. }
\end{quote}

\begin{figure}[t!]
\includegraphics[width=0.99\linewidth]{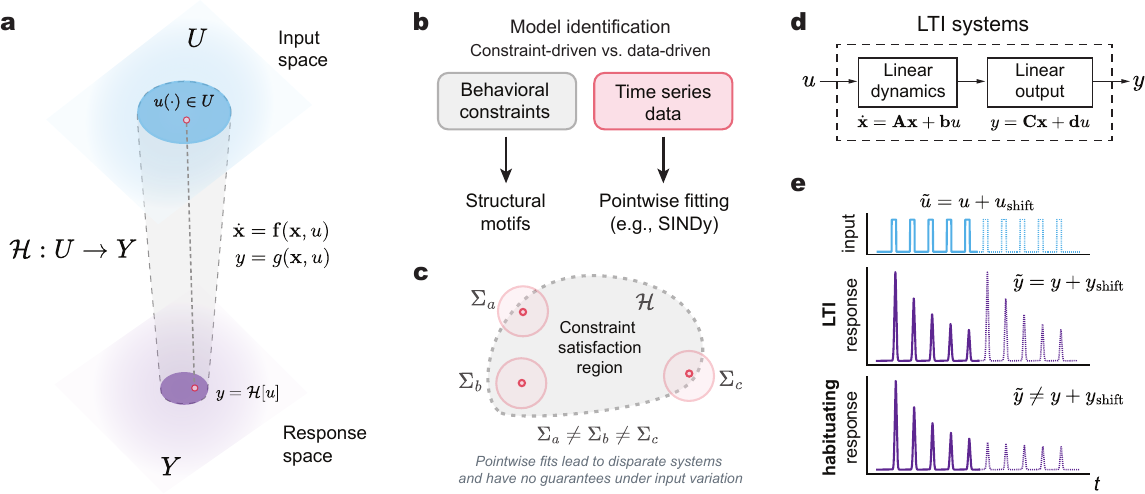}
\caption{
  \textbf{Behavioral framework, constraints on transient response, and LTI insufficiency.}
  (a) A dynamical system $\dot{\mathbf{x}} = \mathbf{f}(\mathbf{x}, u), y=g(\mathbf{x},u)$ with initial condition $\mathbf{x}(t_0)$ induces an input-output operator $\mathcal{H}:U \rightarrow Y$ mapping a space of admissible input signals $U$ to a space of response trajectories $Y$. An example input-output map $y(\cdot)= \mathcal{H}[u(\cdot)]$ is shown. 
  Rather than fitting isolated trajectories, habituation is formulated here as a constraint on a family of admissible input-output behaviors under $\mathcal{H}$, not on isolated trajectories.
  (b,c) Behavioral constraints define a region of structurally admissible input-output pairs satisfying qualitative hallmarks, whereas pointwise system-identification methods fit individual instances from time-series data. The latter need not guarantee structural satisfaction across classes of inputs.
  (d,e) Linear time-invariant (LTI) systems cannot satisfy habituation constraints based on monotone attenuation. By superposition and time invariance, shifting and summing pulse-train inputs yields shifted and summed outputs in an LTI system. Habituating responses violate this property, making nonlinearity necessary (see Sec.~\ref{sec:lti_barrier}). Panels (d,e) adapted from Smart et al.~\citep{smart2024pnas} with permission (CC BY-NC-ND 4.0).
}
\label{fig:fig2_behavioral_and_lti}
\end{figure}

\noindent 
These goals can at least partially be addressed by constructing a 
parametric \emph{model class} whose members, by virtue of their structure, exhibit the habituation hallmarks in a robust fashion, i.e., across a range of their parameters and initial conditions and across families of admissible inputs. This approach only partially meets the goals, however, since the identified model class may not be unique, and because a model structure is assumed a priori. 
We use single-input single-output time-invariant systems of the form
$\dot{\mathbf{x}} = \mathbf{f}(\mathbf{x}, u)$, $y = g(\mathbf{x}, u)$ as the starting point, 
which proved to be versatile enough to represent a broad variety of recurring behaviors in biological signaling processes (see, e.g., \cite{Ferrell201662,Tyson2003-SniffersBuzzersEtc,Briat201615}).
Any system of this form induces an input--output operator $\mathcal{H}: U \to Y$ mapping a space of admissible input signals $U$ to a space of response trajectories $Y$ (\textbf{Figure~\ref{fig:fig2_behavioral_and_lti}a}); behavioral constraints are then properties of $\mathcal{H}$ across the input family, not of isolated trajectories (see the sidebar titled Behavioral Constraints in the Willems Sense).

Throughout the review, we adopt two standing assumptions: 
\textbf{Assumption 1 (SISO).} We focus primarily on the single-input single-output case, $u \in \mathbb{R}$ and $y \in \mathbb{R}$. Although the multi-input hallmarks ($H_{7\text{--}9}$ in Table~\ref{tab:hallmarks-main}) are largely outside our scope, we briefly discuss them and other multivariable extensions in Section~\ref{sec:families}.
\textbf{Assumption 2 (rest state).} The system relaxes toward a unique steady state in the absence of stimulation, so that the operator $\mathcal{H}$ is well-defined relative to a canonical initial condition and the influence of the remote past fades. This assumption is not essential and is violated by multistability and in singular limits of certain adaptive motifs discussed later.

Just like system identification solves a related but different problem, 
several other techniques address related tasks but do not address 
the problem posed here. 
Data-driven model-discovery methods such as sparse regression (SINDy) \citep{brunton2016discovering} and symbolic regression \citep{schmidt2009distilling} essentially extend system identification to discovering candidate models within model families. Just like classical identification methods, they proceed by scoring candidates against particular input-output trajectories. Such methods are well-suited to their intended problem; they simply answer a different question. Given a habituating time series from a particular system, they may return a model that reproduces it. Such a fit remains to be ``pointwise" in the behavioral space in that it finds a model that reproduces a particular input-output behavior rather than discovering the core model features required for qualitative input-output behaviors (\textbf{Figure~\ref{fig:fig2_behavioral_and_lti}b,c}). 
Notably, Volterra series can be used to construct a model for given input and output time data in the same ``pointwise" sense. Volterra series are worth mentioning for their ability to implement a fading memory, which turns out to be one of the recurring features of a habituating system (see Section~\ref{sec:fading-memory}).  

The problem we pose here resembles the search for a \emph{normal form} in bifurcation theory~\cite{GuckenheimerHolmes2002}: 
a representative of minimal complexity that captures a prescribed qualitative behavior, cannot be reduced further, and to which more elaborate systems exhibiting the same behavior can be systematically related (e.g., by Lyapunov-Schmidt reduction~\cite{golubitsky1985singularities}). 
We add one consideration specific to the biological setting: the identified structure should admit a biological interpretation, with terms expressible, for instance, through mass-action kinetics or neural circuit realizations, so that the desired motif is not only mathematically minimal but also mechanistically plausible.

Habituation is one instance of this more general problem. The hallmarks formalized in \textbf{Table~\ref{tab:hallmarks-main}} are the specific qualitative constraints we use throughout this review, but the methodological move, i.e., identifying model classes from behavioral specifications rather than fitting models to data, has a much broader scope and motivates further study.

\begin{textbox}[h]
\section{BEHAVIORAL CONSTRAINTS IN THE WILLEMS SENSE}

The framing of habituation as a constraint on input--output behavior rather than on individual trajectories has a natural grounding in the \emph{behavioral approach} to systems theory \citep{PoldermanWillems1998, willems2007behavioral}. There, a system is defined not by state-space equations but by its \emph{behavior} $\mathcal{B}$: the set of all signal trajectories $(u, y)$ it admits. Systems with different internal representations but identical $\mathcal{B}$ are behaviorally equivalent; the behavior, not any realization, is primary. This is a natural setting for our constraints: taking $\mathcal{B}$ rather than a particular model as the object expresses what the hallmarks restrict: which input--output behaviors are admissible.

Our notion of \emph{behavioral constraint} maps onto this framework. The constraint-satisfaction region in \textbf{Figure~\ref{fig:fig2_behavioral_and_lti}c} can be written as the set of systems whose behaviors contain only trajectories consistent with specific hallmarks across the admissible input family,
\begin{equation*}
  \mathcal{C} = \{ \Sigma : (u, y) \in \mathcal{B}(\Sigma) \text{ satisfies } H_1, H_2, \ldots \text{ for all admissible } u \}.
\end{equation*}
Constraint-driven identification asks for $\mathcal{C}$ itself; data-driven identification returns specific $\Sigma_i$ whose behaviors $\mathcal{B}(\Sigma_i)$ happen to include the observed time-series in a pointwise manner.
The behavioral framework provides a natural language for expressing the hallmarks, but the qualitative constraints they impose (e.g., inequality and ordering relations on responses across admissible input families) differ from the properties it has classically been used to study.
Whether existing behavioral tools can be extended in this direction is, to our knowledge, an open and interesting question.
In that direction, we note that related perspectives have been pursued in the study of control and engineering of excitable systems \cite{sepulchre2018excitable, ribar2021neuromorphic}.
\end{textbox}

\subsection{Hallmarks as constraints on dynamical behavior}

\textbf{Table~\ref{tab:hallmarks-main}} recalls the hallmarks of habituation from Rankin et al. \cite{rankin2009} (which refined the original hallmarks from Thompson \& Spencer \citep{Thompson1966}), together with mathematical formulation proposed in \citep{smart2024pnas}. The criteria translate verbal descriptions (\emph{``progressive decrease", ``spontaneous recovery", ``more frequent stimulation results in more rapid decrement"} etc.) into precise statements about sequences of pulse responses $y[k]$ generated by a periodic input $u(t)$.
We model stimuli as non-negative periodic pulse trains of period $T$, amplitude $A$,
and duty cycle $d\in(0,1)$, and summarize the continuous response $y(t)=g(\mathbf{x},u)$ by its
peak within each period: $y[k] \coloneqq \max_{t\in[kT,(k+1)T)} y(t)$, $k=0,1,\dots$.
The hallmarks then become constraints on the sequence $\{y[k]\}$ rather than on the
full trajectory.

We present and study the hallmarks as \emph{examples} of behavioral constraints on transients. They are derived from decades of study of biological phenomena in the context of habituation, and even there, they should be interpreted cautiously rather than as a definitive checklist. Specifically, only hallmarks $H_1$ (habituation itself) and $H_2$ (spontaneous recovery) are necessary for a system to be said to habituate at all; the remaining hallmarks describe additional structure that some habituating systems exhibit and others do not. This review focuses on $H_1$, $H_2$, and the structurally distinct extensions required for $H_4$ (frequency sensitivity) and $H_5$ (intensity sensitivity); $H_3$, $H_6$, $H_7$, $H_8$ are then addressed via minor modifications of the motif (Section~\ref{sec:families}), while $H_9$ and $H_{10}$ are not focused on here. Furthermore, the passage from verbal descriptions to mathematical criteria is lossy and may admit several interpretations (e.g., see ``and/or" wording in several hallmarks) or alternative formalizations (see, for instance, the numerical filtering criteria for identifying habituating trajectories by parameter search in Methods of Eckert et al. \cite{eckert2024}).

Beyond the ten hallmarks, we introduce one further constraint that we treat as a standing physical assumption rather than a numbered hallmark:

\begin{quote}
\emph{$H_0$ (nonnegativity and boundedness):} the system output satisfies $y(t) \geq 0$ and remains bounded for all admissible inputs.
\end{quote}
\noindent This reflects the fact that biological response variables such as firing rates and concentrations of secreted molecules are physically bounded below by zero and above by saturation. $H_0$ is not specific to habituation; it constrains transients in any setting where the output is a physical quantity. We flag it separately because, as we show next, it underlies the structural argument against purely linear systems.


\begin{table*}[ht]
    \noindent
    \caption{Hallmarks of habituation~\cite{Thompson1966, rankin2009} in shortened form, alongside candidate mathematical criteria$^{\rm a}$. 
    Extended from Table~1 of Smart et al.~\cite{smart2024pnas} CC BY-NC-ND
    }
    \tableshift
    \begin{tabularx}{1.4\linewidth}{l X X}

        \hline
        \textbf{Hallmark} & \textbf{Description} & \textbf{Mathematical criterion} \\
        \hline

        \hline
        $H_1$
        & 
        \footnotesize
        \begin{minipage}[t]{\linewidth}
          \textbf{Habituation}: 
          Repeated application of a stimulus results in a progressive decrease of a response to an asymptotic level. 
        \end{minipage}
        & 
        \footnotesize
        There exists a periodic stimulus $u(t)$ that generates a sequence of responses satisfying 
        $0 \le y[k+1] \le y[k]$ for all $k\ge 0$, and there exists a $K>0$ such that $y[k+1] < y[k]$ for all $k = 0, \ldots, K$. 
        \\

        \hline
        $H_2$
        & 
        \footnotesize
        \begin{minipage}[t]{\linewidth}
          \textbf{Spontaneous recovery}$^{\rm b}$:
          If the stimulus is withheld, the response recovers over time.
        \end{minipage}
        & 
        \footnotesize
        Assume the stimulus has been applied $k$ times, resulting in a response $y[k]$. 
        There exists an $m\in\mathbb{N}$ such that the response after withholding the stimulus for $m$ periods satisfies $y[k+m+1] > y[k]$.
        \\

        \hline
        $H_3$
        & 
        \footnotesize
        \begin{minipage}[t]{\linewidth}
          \textbf{Potentiation of habituation}:
          After multiple series of stimulus repetitions and spontaneous recoveries, the response decrement becomes
          successively more rapid and/or 
          more pronounced. 
        \end{minipage}
        & 
        \footnotesize
        Assume the stimulus is applied for $L$ periods, subsequently withheld $L^\prime$ periods, and assume this pulse-then-rest pattern is repeated. There exists a number $K\in\mathbb{N}$ 
        of pulse-then-rest repetitions 
        such that 
        $y[K(L+L^\prime)+k]< y[k]$ for some subsequent stimuli  $k= 0, \dots, m-1$, $m>0$. 
        \\

        \hline
        $H_4$
        & 
        \footnotesize
        \begin{minipage}[t]{\linewidth}
          \textbf{Frequency sensitivity}:
          Other things being equal, more frequent stimulation results in 
          (a) more rapid and/or more pronounced response decrement, and 
          (b) more rapid recovery.
        \end{minipage}
        & 
        \footnotesize
        Define 
        $U_T = \{ \textrm{stimulus period\,\,} T\in \mathbb{R}^+ |\, \textrm{$H_1$ holds} \}$.
        If $T_1 < T_2 \in U_T$, then 
        (a) the responses satisfy $y_{1}[k] \le y_{2}[k] \ \forall \ k\in \mathbb{N}$, 
        and 
        (b) $m_1 T_1 \le m_2 T_2$ where $m_i T_i$ defines the time to recover following $k$ stimulations
        (i.e. smallest $m_i$ that satisfies $|y_i[0] - y_i[k + m_i]| < \epsilon$). 
        \\

        \hline
        $H_5$
        & 
        \footnotesize
        \begin{minipage}[t]{\linewidth}
          \textbf{Intensity (amplitude) sensitivity}:
          Within a stimulus modality, less intense stimuli give more rapid and/or more pronounced response decrement. 
          Intense stimuli may yield no significant observable response decrement.
        \end{minipage}
        & 
        \footnotesize
        Define $U_A = \{ \textrm{stimulus intensity\,\,} A\in \mathbb{R}^+ \,|\, \textrm{$H_1$ holds} \}$.
        If $A_1 < A_2 \in U_A$,
        then the normalized responses$^{\,\rm c}$ satisfy $y_{1}[k] \le y_{2}[k] \ \forall \ k\in \mathbb{N}$.
        $U_A$ may be bounded above.
        \\

        \hline
        $H_6$
        & 
        \footnotesize
        \begin{minipage}[t]{\linewidth}
          \textbf{Subliminal accumulation}:
          The effects of repeated stimulation may continue to accumulate even after the response has reached an asymptotic level. 
          Among other effects, this can delay the onset of spontaneous recovery.
        \end{minipage}
        & 
        \footnotesize
        Suppose the response reaches an asymptotic level on the $k^{th}$ stimulation. 
        Let $l_1 < l_2 \in \mathbb{N}$. 
        Then a system receiving $k+l_2$ stimulations will take longer to recover than a system receiving $k+l_1$ stimulations.
        \\

        \hline
        $H_7$
        & 
        \footnotesize
        \begin{minipage}[t]{\linewidth}
          \textbf{Stimulus specificity}:
          The response decrement shows some stimulus specificity, [...] distinguishing it from more general sensory adaptation or motor fatigue.
        \end{minipage}
        & 
        \footnotesize
        There exists a second stimulus $s(t)$ such that, when substituted for the original stimulus $u(t)$ during $kT < t < (k+1)T$, the resulting response satisfies $y[k+1] > y[k]$.        
        \\
        
        \hline
        $H_8$
        & 
        \footnotesize
        \begin{minipage}[t]{\linewidth}
          \textbf{Dishabituation}:
          Presentation of another (usually strong) stimulus results in recovery of the habituated response.
        \end{minipage}
        & 
        \footnotesize
        There exists a second stimulus $v(t)$ such that, when presented during $kT < t < (k+1)T$, the response to the next pulse of the original stimulus $u(t)$ satisfies $y[k+1] > y[k]$.
        \\

        \hline
        $H_9$
        & 
        \footnotesize
        \begin{minipage}[t]{\linewidth}
          \textbf{Habituation of dishabituation}: Upon repeated application of the dishabituating stimulus, the amount of dishabituation produced decreases.
        \end{minipage}
        & 
        \footnotesize
        Not considered. 
        \\

        \hline
        $H_{10}$
        & 
        \footnotesize
        \begin{minipage}[t]{\linewidth}
          \textbf{Long-term habituation}:
          Some stimulus protocols may result in properties of the response decrement that last hours, days, or weeks.
        \end{minipage}
        & 
        \footnotesize
        Not considered.
        \\
        
        \hline
        
    \end{tabularx}
    \label{tab:hallmarks-main}
\begin{tabnote}
$^{\rm a}$Hallmarks after $H_1$ assume there exists a stimulus $u(t)$ for which $H_1$ holds. \\
$^{\rm b}$Only $H_1$ and $H_2$ are necessary for a system to be said to habituate. \\
$^{\rm c}$In cases where raw response magnitude grows trivially with amplitude, $H_5$ may be assessed on a normalized response $\tilde y[k] := y[k]/y[0]$ (or via a normalizing output gate, see e.g. Eq.~(5) in Ref. \cite{smart2024pnas}).
\end{tabnote}
\end{table*}

\subsection{The LTI barrier: nonlinearity is necessary}
\label{sec:lti_barrier}

The simplest candidate dynamical structure --- linear time-invariant (LTI) dynamics with a linear output --- is structurally incompatible with habituation as defined above. The argument is short and worth stating explicitly because it sets up the rest of the review: any system that habituates with non-negative output (i.e., satisfies $H_0$ together with $H_1$) must be nonlinear, not as a modeling preference but as a structural requirement.

\emph{Sketch of the argument (see Smart et al.\ \citep{Monnigmann2024CDC}, Prop.~3.1):} 
Consider a finite pulse train $u(t) \ge 0$ consisting of $N$ pulses with period $T$, with $u(t) = 0$ for $t < 0$.
Suppose, for contradiction, that an LTI system $\Sigma$ with input $u(t) \geq 0$ produces a habituating response $y(t) \geq 0$ to $u$, in the sense of $H_1$. 
By time invariance, the response to the time-shifted input $u_{\text{shift}}(t) := u(t - \tau)$ is $y_{\text{shift}}(t) = y(t - \tau)$. 
By superposition, the response to the summed input $\tilde u(t) := u(t) + u_{\text{shift}}(t)$ is the summed response $\tilde y(t) = y(t) + y_{\text{shift}}(t)$. 
However, habituation demands that the response to repeated stimuli attenuate with stimulus history: the peaks of $\tilde y$ should fall below the corresponding shifted-and-summed peaks of $y + y_{\text{shift}}$, since the second pulse train is preceded by the first. This contradicts the superposition identity. 
Hence no LTI system can simultaneously satisfy $H_0$ and $H_1$.
\textbf{Figure~\ref{fig:fig2_behavioral_and_lti}d,e} depicts the argument graphically.

Two points are worth noting. First, the result does not depend on the specific form of the pulse train, only on periodicity and the nonnegativity of $u$ and $y$. Second, the argument generalizes: it applies to any system whose input--output behavior satisfies superposition and time invariance, including infinite-dimensional and distributed LTI systems \citep{Monnigmann2024CDC}. Consequently, habituation with non-negative response requires nonlinearity somewhere in the system --- in the dynamics, in the output map, or both. We address this later in Section~\ref{sec:minimal-motif}.

\section{ADAPTATION VS.\ HABITUATION}
\label{sec:adapt-vs-hab}
Adaptation and habituation are often discussed together, and a recurring question is whether the latter is simply a special case of the former. The two are related but logically independent: adaptation constrains \emph{asymptotic} behavior, whereas habituation constrains the \emph{transient} response.

\subsection{The distinction}

A system \emph{adapts} to a class of inputs $U$ if its output asymptotically converges to a prescribed adapted state for every $u(\cdot)\in U$; the canonical case is $U = \{\text{step functions}\}$, in which case $y(t) \to y^*$ independent of step amplitude. A system \emph{habituates} if, in response to a periodic pulsatile input, the sequence of pulse responses $\{y[k]\}$ satisfies the monotone attenuation and recovery requirements $H_1$ and $H_2$ of Table~1. Adaptation thus constrains $\lim_{t\to\infty} y(t)$; habituation constrains the shape of the response over a sequence of discrete events. 

Eckert et al.~\cite{eckert2024} show that classic adaptive biochemical motifs can exhibit habituation, and Figure~2 of Smart et al.~\cite{smart2024pnas} demonstrates several (sniffer, negative feedback, antithetic integral control; \cite{Tyson2003-SniffersBuzzersEtc, Ma2009, Ferrell201662, Briat201615}) that both adapt and habituate for the same choice of parameters. While this illustrates their close relationship, it does not imply that one property entails the other. To that end, we provide counterexamples in both directions below. 

We also note that while adaptation is often introduced as an asymptotic property \cite{yi2000robust}, work in bacterial chemotaxis has long emphasized that adapting systems can also exhibit rich transient dynamics
\cite{tu2013quantitative} (see also \cite{tu2018adaptation}). The distinction we emphasize is not that adaptation strictly ignores transients but that the habituation hallmarks \emph{constrain} them: they specify which transient shapes are admissible across a family of pulsatile inputs, while adaptation as a design specification leaves substantial freedom in the transient response.

Thus, throughout this section, ``adaptation'' refers specifically to \emph{perfect adaptation} 
in the sense of rejection of step inputs, following standard usage in systems biology~\cite{Ferrell201662,Briat201615,Ma2009,Khammash2019}.
The broader biological usage --- substantial but not necessarily complete attenuation of the response --- is a strictly weaker condition; habituation generally implies it, since response amplitudes decrease across repeated pulses. The interesting logical question, and the one we address below, is whether habituation implies \emph{perfect} adaptation.

\subsection{Adaptation does not imply habituation}

We give two examples. A damped harmonic oscillator separates adaptation from habituation transparently but only by dropping $H_0$ (its output changes sign). We then show that a nonlinear motif (AIC) exhibits the same separation while respecting $H_0$ --- reinforcing the positivity assumption of the LTI barrier in Section~\ref{sec:lti_barrier}. 

For the damped harmonic oscillator, take velocity as the response variable under input forcing. A unit mass with position $q(t)$ subject to a linear restoring force, frictional damping, and an external force, has the dynamics:
\begin{align}
\label{eq:damped_oscillator}
    \ddot{q} + \gamma\,\dot{q} + k\,q \,=\, u(t), \qquad k, \gamma > 0,
\end{align}
where $k$ is the stiffness and $\gamma$ the damping coefficient. 
In state-space form with $x_1=q$, $x_2=\dot q$ and output $y=x_2$, the system can be interpreted as a first-order lag subject to integral feedback $\dot{y}+ \gamma y= u- kx$, $\dot{x}= y$ with controller gain $k$.
Two features of this system are apparent.

\emph{Adaptation is structural.} 
Under constant input $u_0$, the mass settles at $q^*=u_0/k$ with $\dot q\to 0$ for any $\gamma>0$. 
Since $\dot x_1 = x_2 = y$, 
the position integrates the output over time and stores the offset required for the spring force $k x_1$ to balance the constant disturbance, asymptotically driving $y \to 0$ for any $u_0$.
This linear integrator thus acts as an internal model for constant inputs in the sense of the internal model principle (IMP)~\cite{FrancisWonham1976_IMP,sontag2022_annurev_internalmodels}. 

\begin{figure}[t!]
\includegraphics[width=0.99\linewidth]{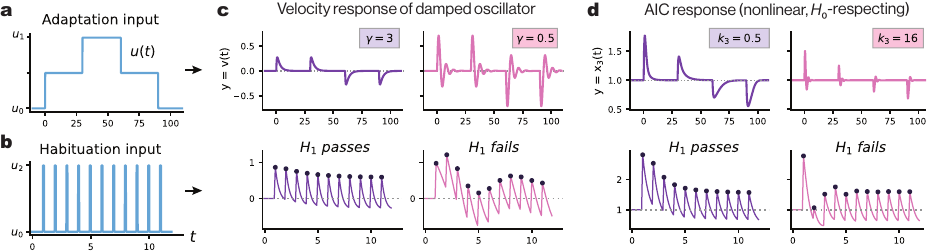}
\caption{
  \textbf{Perfect adaptation does not imply habituation.} 
  Two systems that adapt to a step input but differ in whether they habituate to a pulse train. 
  (a)~Adaptation input: a staircase of amplitudes $u_0 \to u_1 = u_0 + 2A$ (steps $A=1$; the oscillator uses baseline $u_0=0$, the AIC $u_0=1$). 
  (b)~Habituation input: a pulse train of period $T=1$ and duty $d=0.1$ reaching $u_2 = u_0 + A/d$, chosen so the integrated input per period matches the step.
  (c)~Damped oscillator, Eq.~\eqref{eq:damped_oscillator} (linear; velocity output, so $H_0$ is dropped). Top: velocity adapts to each staircase level for both damping values ($k=1$; $\gamma=3$ overdamped, $\gamma=0.5$ underdamped). Bottom: the overdamped peak sequence satisfies $H_1$; the underdamped one violates it. 
  (d)~AIC motif, Eq.~\eqref{eq:AIC} (nonlinear; output $x_3\ge 0$). 
  Top: $x_3$ adapts to the same level $x_3^*=k_1/k_2$ for both parameter choices. Bottom: the pulse-peak sequence satisfies $H_1$ for $k_3=0.5$ and violates it for $k_3=16$ ($k_1=k_2=k_4=k_5=2$).
}
\label{fig:fig3_adapt_not_hab}
\end{figure}

\emph{The transient depends on damping.} 
With damping ratio $\zeta = \gamma/(2\sqrt{k})$, the overdamped case ($\zeta \geq 1$, weak feedback gain) gives a monotone pulse-peak sequence that robustly satisfies $H_1$, while the underdamped case ($\zeta < 1$, strong feedback gain) exhibits damped oscillations, and for suitable periods the peak sequence fails to decrease monotonically, violating $H_1$ (\textbf{Figure~\ref{fig:fig3_adapt_not_hab}c}). The internal model guarantees adaptation independently of $\zeta$, while the damping regime affects habituation.

Separating the two \emph{within} the $H_0$-respecting class requires nonlinearity, as illustrated by antithetic integral control (AIC) --- a canonical motif for perfect adaptation~\cite{Briat201615, Khammash2019}:
\begin{equation}
\label{eq:AIC}
    \begin{split}
        \dot{x}_1 &= k_1 - k_3 x_1 x_2, \\
        \dot{x}_2 &=  - k_3 x_1 x_2+ k_2 x_3, \\
        \dot{x}_3 &= - k_5 x_3+ k_4 x_1 u , 
    \end{split}
\end{equation}
with output $y= x_3$, and $k_i>0$. 
The system has a unique steady state for all constant positive inputs $u_0$, 
with output $y^\ast = k_1/k_2$ independent of $u_0$.
This steady state is the only attractor in the positive orthant, which is positively invariant for positive inputs and parameters.
These properties imply perfect adaptation to positive step inputs. 
While the AIC motif adapts \emph{by design}, it violates hallmark $H_1$ for some parameter values while respecting $H_1$ for others, as \textbf{Figure~\ref{fig:fig3_adapt_not_hab}d} demonstrates. The two behaviors, habituating and non-habituating, persist under finite variations of parameters, duty, period, and amplitude of the pulse train at their input, while perfect adaptation holds throughout. 

Both habituation and adaptation are nonlinear phenomena under positivity. As an aside, the LTI barrier raises the natural question of whether perfect adaptation with $H_0$ is possible for a \emph{positive} LTI system~\cite{Farina2000}: i.e., a system of the form $\dot{\mathbf{x}} = \mathbf{A}\mathbf{x} + \mathbf{b}u$, $y = \mathbf{c}^T \mathbf{x}$ with $x_i(t)\ge 0$ for all components and $y(t)\ge 0$ for all nonnegative initial conditions and nonnegative inputs. No such system exists, however, which can be seen with elementary arguments. First note that the impulse response $h(t)= \mathbf{c}^T e^{\mathbf{A}t} \mathbf{b}$ of a positive LTI system is nonnegative~\cite[Chapter 2]{Farina2000}. Perfect adaptation requires zero DC gain $G(0)= \int_0^\infty h(t)\,dt$, which implies $h(t)=0$ for all $t$ if $h(t)\ge 0$ for all $t$. An impulse response that is identically zero, however, implies the output $y(t)= \int_0^t h(t-\tau) u(\tau)\, d\tau$ is zero for all times and any input, or equivalently, $G(s)= 0$ for all $s$. In other words, the trivial system with $G(s)=0$ is the only positive LTI system that achieves perfect adaptation.

Ref.~\cite{Monnigmann2024CDC} showed that no positive LTI system exists that habituates (or, more precisely, no such system implements $H_0$, $H_1$, and $H_2$, which implies the former statement). Essentially, both adaptation and habituation are intrinsically nonlinear under positivity. 
The harmonic oscillator above evades this only by dropping $H_0$: with velocity as output it \textit{adapts and habituates} when overdamped and \textit{adapts but does not habituate} when underdamped, but its sign-changing output places it outside the positive class. The AIC motif serves as a canonical nonlinear counterexample: it structurally respects $H_0$ and shows that perfect adaptation (to step inputs) does not imply habituation.

\subsection{Habituation does not imply adaptation}
\label{subsec:habi-does-not-imply-adap}

The converse direction also fails. The minimal habituation motif we introduce next in Sec.~\ref{sec:families} satisfies $H_1$ and $H_2$ in a broad region of parameter space, but its asymptotic response to a sustained constant input is generically nonzero. Habituation gates the \emph{transient} response to repeated pulses; it does not require the asymptotic response to a step to vanish. The same motif spans both regimes depending on parameters --- the steady-state attenuation can be made small in appropriate limits --- but a nonzero asymptote is the generic case.

This is also consistent with the biological motivation for habituation: it often acts as a fast, imprecise filter rather than an exact asymptotic rejector. Organisms in fluctuating environments cannot wait arbitrarily long for an asymptote, and partial suppression may itself be advantageous, as has been suggested experimentally (see Section~\ref{sec:inference_biology} for one example).

A natural question is whether the converse can be recovered under stronger hypotheses. E.g., if $H_1$ held uniformly across pulse trains of all periods $T$, then in the limit $T\to0$ (or duty $d\to1$) the input approaches a constant step, suggesting a heuristic link between habituation and adaptation via averaging arguments. The required assumptions appear substantially stronger than those relevant biologically; we leave this as an open direction.

Adaptation and habituation therefore depend on distinct features of the dynamics: the internal model determines asymptotic rejection, whereas the parameter regime governs whether transient responses exhibit habituation.

\section{STRUCTURAL FAMILIES OF HABITUATING SYSTEMS}
\label{sec:families}

The structural argument of Sec.~\ref{sec:formalize} establishes that habituation requires nonlinearity. 
Known adaptive systems can habituate for suitable parameters, as shown in Smart et al.~\cite{smart2024pnas}, but the converse fails: Sec.~\ref{sec:adapt-vs-hab} establishes that adaptation does not imply habituation, and is not the lens we use to find one. The first result bounds the space of admissible motifs from below; the second means asymptotic adaptation offers no shortcut to identifying a motif that is \emph{minimal} in dimensionality, in nonlinearity, and in parameter count. In this section, we construct such a motif directly from the hallmarks, treating them as design specifications as in that work, and then situate the resulting structure within the broader literature.

\subsection{Constructing a minimal motif from the core hallmarks}
\label{sec:minimal-motif}

Consider the goal of building the simplest dynamical system that satisfies $H_0$ (nonnegative bounded output), $H_1$ (progressive response decrement), and $H_2$ (spontaneous recovery). The construction below, adapted from \citep{smart2024pnas}, proceeds in three steps, each motivated by a shortfall of the one before. 
While we adopt a multiplicative filter below, the exact form of the static nonlinearity is not critical: any function that passes low-memory input, attenuates high-memory input, and keeps the output nonnegative will suffice.

\emph{Step 1: A multiplicative receptivity filter.}  
Rather than starting from the dynamics, start from the desired output, writing it as
\begin{equation}
  y(t) = u(t)\,\sigma(t),
\end{equation}
where $u(t) \geq 0$ is the input and $\sigma(t) \in [0, 1]$ is a time-varying ``receptivity'' that gates the response. The simplest nontrivial choice is an exponential envelope $\sigma(t) = e^{-\alpha t}\,H(t)$, with $H$ the Heaviside step. This trivially satisfies $H_1$ --- the response decreases monotonically with each successive pulse --- but fails $H_2$, because the receptivity decays to zero regardless of whether the stimulus continues or is withheld. Recovery requires that $\sigma$ depend on the recent input history, not on an absolute time that privileges $t=0$.

\emph{Step 2: Fading memory of the input.} 
Replace the fixed envelope with a memory of recent input, the simplest choice being a first-order leaky integrator or lag:
\begin{equation}\label{eq:minimal-motif-x}
\dot{x} = \beta u - \alpha x,
\qquad
x(t) = \beta\!\int_{-\infty}^{t} e^{-\alpha(t-\tau)}\,u(\tau)\,d\tau ,
\end{equation}
where $x(-\infty)=0$.
Intuitively, $x(t)$ tracks a weighted average of $u$ over the recent past, with the weighting determined by $\alpha$. When the stimulus is sustained, $x$ rises; when it is withheld, $x$ decays. Equation~(\ref{eq:minimal-motif-x}) is the minimal continuous-time fading-memory unit: a single state variable, two parameters, linear dynamics.
\footnote{Equivalently, the internal state is a convolution, $x = h * u$ with $h(t) = \beta e^{-\alpha t} H(t)$ --- the simplest fading-memory filter, a single linear convolution of the input. More general nonlinear fading-memory operators can be represented through Volterra series, which augment this first-order kernel with higher-order interactions among past inputs (see Section~\ref{sec:fading-memory}; cf.~\citealp{boyd1984analytical,maass_and_sontag2000_neural_as_nonlin_filter}).}

\begin{figure}[t!]
\includegraphics[width=0.89\linewidth]{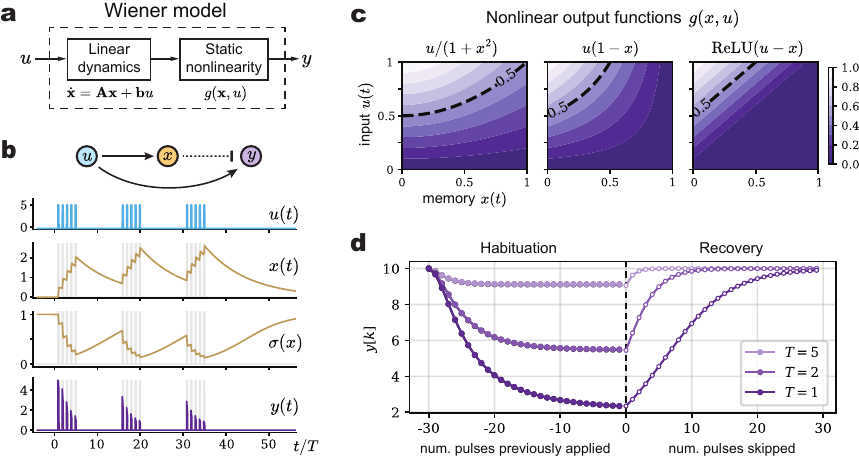}
\caption{
  \textbf{Minimal motif construction and properties.}
  (a) Wiener block diagram: linear dynamics followed by a static nonlinearity.
  (b) Mechanism walkthrough showing time-traces of input $u(t)$, memory state $x(t)$, receptivity $\sigma(x)$, and output $y(t)$ for a representative pulse-train stimulus. 
  (c) Three viable output functions $g(x, u)$ producing qualitatively similar habituating responses, illustrating structural robustness.
  (d) Peak response $y[k]$ as a function of pulse number for three stimulation periods, with recovery monitored after 30 pulses (vertical dashed line). 
  Adapted from Smart et al.~\citep{smart2024pnas} with permission (CC BY-NC-ND 4.0).
}
\label{fig:fig4_wienermodel}
\end{figure}

\emph{Step 3: Nonlinear readout.} 
Recalling that habituation requires nonlinearity somewhere in the system (Sec.~\ref{sec:lti_barrier}), and having kept the dynamics linear, we place it in the readout: couple the receptivity to the memory state through a monotone-decreasing function, e.g.,
\begin{equation}
\label{eq:minimal-motif-y}
  y = u\,\sigma(x), \qquad \sigma(x) = \frac{1}{1 + x^N},
\end{equation}
for some $N \geq 1$. 
Note that the receptivity $\sigma(\cdot)$ is now implicitly a function of time through the memory $x(t)$. When $x$ is small (few recent stimuli), $\sigma(x) \approx 1$ and the system passes the input through. When $x$ is large (saturated by recent stimuli), $\sigma(x) \approx 0$ and the response is attenuated. This choice of $\sigma$ is one example; any function with $\sigma(0) = 1$ that is positive and monotone-decreasing over the range of $x$ yields qualitatively similar behavior (\textbf{Fig. \ref{fig:fig4_wienermodel}c}).

Equations~(\ref{eq:minimal-motif-x})--(\ref{eq:minimal-motif-y}) together constitute an instance of the minimal motif: linear dynamics composed with a static nonlinear output. 
This is a \emph{Wiener model} in the block-oriented systems literature \citep{Schoukens2017}: linear dynamics followed by a static nonlinearity (\textbf{Figure~\ref{fig:fig4_wienermodel}a}). The mechanism is transparent in time-traces (\textbf{Figure~\ref{fig:fig4_wienermodel}b}): pulses drive $x$ upward, $\sigma(x)$ correspondingly drops, and the response $y$ attenuates across successive pulses; during quiescent periods, $x$ decays and the system recovers. 
\textbf{Figure~\ref{fig:fig4_wienermodel}d} shows that response peaks decrease monotonically across a range of stimulation periods, and that recovery is complete given sufficient quiescent time.

A multiplicative output nonlinearity is not essential. In the more general case, one has
\begin{equation}
\label{eq:minimal-motif-gen}
  \dot x = \beta u - \alpha x, 
  \qquad 
  y = g(x,u),
\end{equation}
with several examples of attenuating $g(x,u)$ shown in \textbf{Figure \ref{fig:fig4_wienermodel}c}.
E.g., in an analog circuit context, a diode can realize a ReLU-type nonlinearity which mediates the same behavior \cite{Monnigmann2024CDC}. This thresholded subtraction output is related to ``negative image" models discussed in the biology literature \cite{ramaswami2014network, Shen2020_odor_negative_image_model}. 

Finally, we note that the motif admits a singularly perturbed realization, 
\begin{equation}
\label{eq:minimal-motif-singular}
  \dot x = \beta u - \alpha x, 
  \qquad 
  \epsilon\dot y = -y +  g(x,u),
\end{equation}
with the Wiener structure recovered in the limit $\epsilon \rightarrow 0$. Thus, a single dynamic memory variable and a nonlinear readout are sufficient for habituation ($H_1$) and recovery ($H_2$). 

Several features of this construction are notable. The structure was not assumed but \emph{derived} from the hallmarks: each step was motivated by the next requirement, and the resulting system appears difficult to simplify further without violating one of the constraints. The dynamics for $x$ are linear despite the system as a whole being nonlinear; the nonlinearity is confined to the static output map. And the motif is biologically plausible: $x$ can be interpreted as a leaky molecular memory (fixed degradation rate $\alpha$, input-dependent production $\beta u$), and $g(x,u)$ as a saturating regulation of the response (e.g., Hill-like attenuation by phosphorylation \citep{Gunawardena2005}). The motif is thus both mathematically economical and mechanistically interpretable. Its small parameter count also makes it amenable to Bayesian inference of individual-level dynamical phenotypes from behavioral data, an application we return to in Section~\ref{sec:realizations}.

\subsection{Structural extensions for additional hallmarks}
\label{sec:extensions}

The motif satisfies the hallmarks $H_0$ through $H_3$ as well as $H_{4\textrm{(a)}}$. For the latter, observe that \textbf{Figure~\ref{fig:fig4_wienermodel}d} shows that more frequent stimulation produces more rapid and pronounced decrement, measured by response level as a function of stimulus index. The remaining hallmarks require structural additions. Two of these are noteworthy because they correspond to mechanistically distinct ingredients: \emph{series composition} for $H_{4\textrm{(b)}}$ (faster recovery under more frequent stimulation) and \emph{input nonlinearity} for $H_5$ (intensity sensitivity). 

\subsubsection{Series composition}
For the second aspect of frequency sensitivity ($H_{4\textrm{(b)}}$: that more frequent stimuli produce more rapid \emph{recovery}), a single Wiener unit is insufficient. The single-unit recovery time is governed by $\alpha$ alone and does not depend on stimulation frequency. Connecting two Wiener units in series, with separated timescales $\alpha_1 \gg \alpha_2$, resolves this. Similar observations in discrete-time systems were noted by Staddon \cite{Staddon1993, Staddon1996}, an early advocate of parsimonious dynamical models of behavior \cite{staddon2001adaptive}. The first unit acts as a fast leaky filter: under frequent stimulation, $x_1$ saturates and the output $y_1$ remains small, so the second unit's memory $x_2$ accumulates slowly. Under infrequent stimulation, $x_1$ relaxes between pulses, $y_1$ transmits effectively, and $x_2$ accumulates more rapidly. The result is a frequency-dependent recovery rate, satisfying $H_{4\textrm{(b)}}$ (\textbf{Figure~\ref{fig:fig5_hammerstein}a,b}); see \cite{smart2024pnas} for additional details. 
Eckert et al.~\cite{eckert2024} likewise follow Staddon's observation, capturing frequency sensitivity in a biochemical setting by concatenating motifs. While they rationalize the resulting frequency sensitivity through timescale separation between the slow memory variables, we note that a separate timescale separation can substantially reduce the dimensionality of their numerically identified systems, perhaps at the cost of physical interpretability. E.g., the states $I_1, I_2$ in Figs.~2--3 of \cite{eckert2024} appear to act as quasi-static maps of their inputs; adiabatic elimination removes them and their associated rate constants, leaving a lower-dimensional core closer to the models studied here. Whether such concatenated networks reduce fully to normal-form-like motifs is an interesting open question.

\subsubsection{Input nonlinearity}
\begin{figure}[t!]
\includegraphics[width=0.99\linewidth]{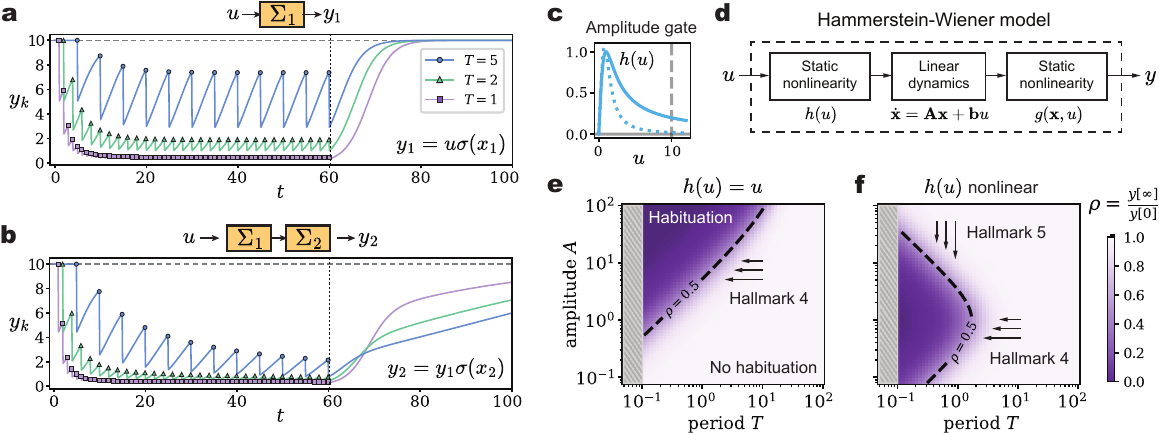}
\caption{
 \textbf{Structural extensions of the minimal Wiener motif: series connection and Hammerstein-Wiener generalization.}
  (a) Response of a single Wiener unit to pulse trains of three different periods $T = 1, 2, 5$; pulses applied until $t = 60$ (vertical dashed line), recovery monitored afterward. The single unit habituates more rapidly for more frequent stimuli but approaches a common recovery asymptote.
  (b) Response of two Wiener units in series with separated timescales. The series configuration acts as a nonlinear low-pass filter: more frequent stimuli saturate the first unit, preventing transmission to the second, and producing frequency-dependent recovery rates.
  (c) Example amplitude gate $h(u) = 2u/(1 + u^N)$ for $N = 2$ (solid) and $N = 3$ (dashed).
  (d) Hammerstein-Wiener motif: static input nonlinearity $h(u)$, linear dynamics, static output nonlinearity $g(\mathbf{x}, u)$.
  (e) Habituation phase diagram for the minimal Wiener motif with $h(u) = u$ and $g(x,u)=\textrm{tanh}(\gamma u)\sigma(x)$, $\gamma \gg 1$, showing asymptotic response ratio $\rho \equiv y[\infty]/y[0]$ across amplitude $A$ and period $T$. Habituation with $H_{4\textrm{(a)}}$ occurs for sufficiently large $A/T$.
  (f) Nonlinear amplitude gate reshapes the habituation region, enabling intensity sensitivity ($H_5$; intense stimuli yield weaker habituation). 
  Adapted from Smart et al.~\citep{smart2024pnas} with permission (CC BY-NC-ND 4.0).
}
\label{fig:fig5_hammerstein}
\end{figure}

Intensity sensitivity ($H_5$) requires the system to respond differently to stimuli of different amplitudes in a manner that the directly attenuating output structure $y = g(x,u)$ alone cannot produce. Prepending a static input nonlinearity $h(u)$ before the linear dynamics introduces this capacity, and corresponds to a Hammerstein-Wiener architecture (\textbf{Figure~\ref{fig:fig5_hammerstein}c,d}). A non-monotone gate such as $h(u) = 2u/(1 + u^N)$ has the effect of compressing strong inputs while passing weak ones, so that strong stimuli drive the memory state less effectively than their raw amplitude would suggest. The asymptotic habituation strength $\rho = y[\infty]/y[0]$ then becomes amplitude-dependent in the manner of $H_5$ (\textbf{Figure~\ref{fig:fig5_hammerstein}e,f}).

\subsubsection{Other hallmarks}
Three further hallmarks are accessible through minor modifications and do not require structural ingredients beyond those above. $H_6$ (\emph{effects of repeated stimulation continue to accumulate after response saturates}; termed ``subliminal accumulation" in Eckert et al.~\cite{eckert2024}) is supported when the output uses an explicit threshold such as $\mathrm{ReLU}(u - x)$, allowing the memory state to continue accumulating after the response has reached zero. 
$H_7$ (stimulus specificity, which rules out fatigue) may be addressed, for instance, by introducing additional input channels feeding parallel Wiener units and taking the maximum of their outputs. $H_8$ (dishabituation by a second stimulus) is supported by adding a depletion term $-\kappa v x$ to Eq.~\eqref{eq:minimal-motif-x}, where $v(t)$ is an additional dishabituating stimulus; when $v$ is presented, $x$ is rapidly depleted and the response recovers. Demonstrations of $H_6$, $H_8$ are given in Fig.~5 of \citep{smart2024pnas}.

\subsection{Related models in the literature}
\label{sec:related-models}
Habituation has been modeled across diverse domains, from neuroscience to systems biology, and we do not attempt a comprehensive review of those literatures; instead, we highlight a recurring structural theme. Across substrates and modeling traditions, habituation is repeatedly implemented through a memory of recent stimulation coupled to a mechanism that attenuates subsequent responses. The models differ in dimensionality, physical interpretation, and derivation, but many can be understood as elaborations of this core motif.

\subsubsection{Neural and behavioral models}
Much of the neuroscience and behavioral literature proposes low-dimensional 
dynamical models in which repeated stimulation modifies an internal response variable whose dynamics govern attenuation and recovery. Examples include Staddon's discrete-time models \citep{Staddon1993,Staddon1996}, along with a broader literature of related models \cite{Stanley1976, DLWang1993, Dragoi2002}; for entry points see \citep{delrosal2006} and references therein. At the cellular level, habituation has long been associated with short-term synaptic depression, studied most influentially in the \emph{Aplysia} gill-withdrawal reflex \citep{kandel1976cellular}. 
In standard models \cite{tsodyks1997neural}, synaptic resources are depleted by activity but recover between stimuli, providing a memory of recent input that attenuates subsequent responses (though circuit-level interactions can produce the same behavior without single-synapse depression). 
More broadly, models of neural activity have employed cascades of linear--nonlinear units \citep{Aljadeff2016,WeberFairhall2019}, closely related to interconnections of the Wiener and Hammerstein--Wiener motifs discussed in Section~\ref{sec:extensions}.

\subsubsection{Novelty detection and negative-image models}
A parallel tradition views habituation as the suppression of familiar input in order to emphasize novelty. Early ideas along these lines arise in Sokolov's work \citep[p. 286-289]{Sokolov1963perception}. Kohonen's novelty filter \citep{textbook_kohonen_1989} provided explicit computational realizations, returning the component of an input that differs from a learned background. In modern sensory neuroscience, similar mechanisms are often discussed in terms of \emph{negative-image} models \citep{ramaswami2014network}, in which inhibitory pathways learn to cancel recurring stimuli. Shen et al.\ \citep{Shen2020_odor_negative_image_model} instantiate this idea in an olfactory setting, modeling habituation as a discrete-time, channel-wise subtraction via a transient inhibitory memory. In our notation, the continuous-time analog of their dynamics takes the form 
$\dot{\mathbf{x}} = \beta \langle \mathbf{u} - \mathbf{x} \rangle_+ - \alpha \mathbf{x} $, 
followed by a static nonlinear output step. Bourassa et al.\ \citep{bourassa2026prxlife} consider the more realistic setting of olfactory circuits exposed to fluctuating odor environments, extending the constant-background setting studied by Shen et al.\ and proposing a background-learning mechanism suited to that regime. These models extend habituation beyond the single-input setting considered here and connect naturally to the multivariable questions in Section~\ref{sec:open}.

\subsubsection{Biochemical and substrate-independent models}
The same attenuate-and-recover structure appears outside neural systems, suggesting that habituation reflects general dynamical principles rather than specifically neural mechanisms. Eckert et al.\ \citep{eckert2024} screened incoherent-feedforward and negative-feedback biochemical circuits for hallmark satisfaction, identifying biochemically plausible networks that exhibit multiple hallmarks simultaneously. While these models are considerably higher-dimensional than the minimal motif and were discovered through numerical search rather than direct construction, they demonstrate that habituation can arise within realistic molecular architectures. Building on this and \cite{smart2024pnas}, Pla-Mauri and Sol\'e \citep{sole2026acsSynthBiol} argue that related circuits could be assembled synthetically from standard genetic components. Bonzanni et al.\ \citep{Bonzanni2019} propose a generalized discrete-time model of habituation that abstracts away from any particular biological substrate. Nicoletti et al. \citep{nicoletti-busiello-2024} analyze a minimal model of habituation involving a slow storage variable that mediates negative feedback, showing that intermediate levels of habituation maximize information gain by balancing information acquisition against energetic dissipation. 

\subsubsection{A methodological observation}
The models above differ not only in substrate but also in how they are obtained. 
The minimal motif of Section~\ref{sec:minimal-motif} was derived analytically from behavioral constraints; Eckert et al.\ \citep{eckert2024} identify habituating systems through numerical search of biochemical circuit space; Gershman \citep{gershman_2024} derives habituation from a normative principle, showing that it emerges from Bayes-optimal filtering in an environment with recurring background stimulation; and Komatsu et al.\ \citep{Komatsu2025} learn habituating input--output operators directly from data using a Fourier neural operator. These approaches provide complementary routes to the same question: what dynamical structures can realize habituation? We return to the broader problem of identifying structure from behavioral specifications in Section~\ref{sec:open}.

\subsection{Fading memory as a shared architectural feature}
\label{sec:fading-memory}

A common feature recurs across the families surveyed above. A leaky memory of recent input coupled to a nonlinear readout is, formally, an instance of \emph{fading memory} (FM) in the sense of Boyd and Chua \cite{BoydChua1985fading}: an input--output operator has FM if the influence of the past on the present output decays over time, so that inputs agreeing on the recent past produce nearly equal responses (see the sidebar on FM for precise statements). Spontaneous recovery ($H_2$) can be viewed as a behavioral expression of this property: the response to a withheld stimulus returns to baseline precisely because the memory of past input fades.
This emphasis on transient responses is consistent with recent discussions of computation in living systems beyond purely attractor-based descriptions \cite{koch2024biological}.
As one related example, memory-kernel formulations have borne fruit in studies of plant tropism, where experimentally inferred kernels suggest that plants can compare recent and earlier stimuli across timescales~\cite{riviere2023plants}.

Fading memory also carries a classical structural consequence: Boyd and Chua \cite{BoydChua1985fading} showed that FM operators admit uniform approximation by Volterra series \cite{schetzen1980volterra, rugh1981nonlinear}, realizable as stable linear dynamics with a static polynomial readout
(recent work recasts FM in a state-space setting via incremental stability notions~\cite{bainier2026state}).
This property contextualizes the Wiener motif within a broader class. In practice, though, the Volterra representation is of limited use for the thresholded, piecewise readouts typical of habituation, e.g., $\mathrm{ReLU}(u-x)$, which may not admit low-order truncations, a difficulty noted more generally \cite[ch. 2.1]{rieke1996spikes}.
Our concern is in any case constructive: identifying minimal motifs from behavioral constraints rather than approximating a given operator. What recurs across the families above is FM itself, not any particular representation of it.

The habituating motif is perhaps the simplest member of this class: a single FM unit with one nonlinear readout; it is also the lowest-dimensional structure that meets the core hallmarks. The same property characterizes far larger systems, reservoir computers and modern state-space models among them, in which high-dimensional FM dynamics support general-purpose computation. We develop that high-dimensional end of the spectrum, and its connections to machine learning, in Section~\ref{sec:ml-connections}.

\begin{textbox}[h]
\section{FADING MEMORY, VOLTERRA SERIES, AND STATE-SPACE REALIZATIONS}
\label{sidebar:fading-memory}


Following Boyd and Chua \citep{BoydChua1985fading}, a time-invariant operator $\mathcal{F}$ mapping inputs on $(-\infty,0]$ to outputs has \emph{fading memory} on a set of bounded inputs if there exists a decreasing weighting function
$w:[0,\infty)\to(0,1]$ with $w(t)\to0$ as $t\to\infty$ 
such that for each $u\in K$ and $\varepsilon>0$ there exists a $\delta>0$ such that for all $v\in K$,
\[
\sup_{t\le0}|u(t)-v(t)|\,w(-t)<\delta
\quad\Rightarrow\quad
|(\mathcal{F}u)(0)-(\mathcal{F}v)(0)|<\varepsilon .
\]
They motivate their definition in the context of early work by Volterra \cite{volterra1959theory} and Wiener \cite{wiener1958nonlinear}; intuitively, FM means that two input signals that are close in the recent past, but not necessarily in the remote past, produce similar present outputs. 

\emph{Volterra series} \cite{boyd1984analytical} generalize the convolution representation of a linear system to nonlinear operators,
\[
y(t)=h_0+\sum_{n\ge1}
\int\!\cdots\!\int
h_n(\tau_1,\ldots,\tau_n)
\prod_{j=1}^{n}u(t-\tau_j)\,
d\tau_j ,
\]
where the kernels $h_n$ encode higher-order nonlinear interactions and $h_1$ is the ordinary impulse response. 

Boyd and Chua showed that, under additional technical assumptions, FM has strong structural consequences. Any time-invariant FM operator can be approximated uniformly, on bounded input sets, by a finite Volterra series. Further, they showed the approximating operator can be realized as a finite-dimensional dynamical system
\[
\dot x = Ax + bu,
\qquad
y = p(x),
\]
where $A$ is exponentially stable and $p$ is a polynomial readout map \citep[Thm.~1--2]{BoydChua1985fading}. Thus fading-memory operators admit approximation by stable linear dynamics coupled to static nonlinear readouts. The habituation motifs discussed in Section~4 may be viewed as low-dimensional members of this broader class.
\end{textbox}

\section{PHYSICAL AND ALGORITHMIC REALIZATIONS} 
\label{sec:realizations}
The minimal motif and its extensions are realized in three distinct ways: as physical devices (electronic circuits, neuromorphic materials, synthetic genetic circuits), as low-parameter models for behavioral inference, and, in a more abstract sense, as algorithmic primitives in machine learning. We survey each in turn, noting that the common thread of transient, fading memory dynamics coupled to a nonlinear readout underlies all three. 

\subsection{Circuits, materials, and synthetic biology}
\label{sec:sec5_engineered}

Closely related properties are realized across several domains 
that differ in physics but share underlying mathematical structure.

\subsubsection{Analog circuits}
Perhaps the simplest hardware implementation is an RC circuit with diode loop (\textbf{Figure~\ref{fig:fig1_overview}b}): the voltage across a capacitor plays the role of the memory variable $x$, and the output voltage, measured across the diode resistor, involves a ReLU-like nonlinearity \cite{Monnigmann2024CDC}. The circuit's rate constants are set by component values, the recovery hallmark $H_2$ follows from capacitor discharge, and habituation under pulsatile input voltage can be directly measured in the voltage measured across the diode resistor. The construction makes the role of each component transparent: the capacitor is the fading memory, the diode gives a static threshold nonlinearity, and $T_{RC}$ provides the recovery timescale.

\subsubsection{Parallels with memristive systems}
Memristive systems provide a useful structural and methodological parallel to the framework developed here, rather than constituting a specific model of habituation. The general memristive system of Chua and Kang \cite{Chua1976-memristiveSystems} takes the form
$\dot{\mathbf{x}} = f(\mathbf{x},u,t)$,
$y = g(\mathbf{x},u,t)\,u$,
so that the instantaneous output is the input scaled by a state-dependent gain. 
The Wiener motif of Section~\ref{sec:minimal-motif} with multiplicative nonlinearity $y=u\, \sigma(x)$ falls naturally within this class.
A useful contrast is with the ideal memristor \cite{chua1971memristor}, whose state variable is charge $q$ satisfying $\dot q=i$ and whose memristance depends only on that accumulated charge. Habituation differs in requiring a leaky memory that eventually forgets past stimulation. The ideal memristor therefore does not natively exhibit spontaneous recovery ($H_2$), although relaxation can be incorporated within the broader memristive framework introduced in~\cite{Chua1976-memristiveSystems}. 
Notably, memristive systems are characterized through qualitative input--output signatures \cite{Chua1976-memristiveSystems}, in a way that parallels the behavioral-constraints perspective developed here for habituation hallmarks (see the memristor sidebar).
And more broadly, memristive ideas have been invoked in models of learning-like behavior in aneural living systems such as \emph{Physarum}  \citep{Pershin2009_memristive_amoeba}.

\begin{textbox}[h!]
\section{GENERIC MEMRISTOR PROPERTIES AS BEHAVIORAL CONSTRAINTS}
\label{sidebar:memristor}
In \cite{Chua1976-memristiveSystems}, Chua and Kang introduce a broad class of memristive systems together with a list of qualitative input-output signatures proposed for their identification, including pinched input--output curves, unique globally asymptotic steady states, and the inability to discharge energy transiently (in contrast, e.g., to a passive RLC circuit). 
This resembles the perspective adopted in Section~2: characterizing a phenomenon through qualitative properties of its input--output behavior rather than through a specific realization. 

The \emph{pinching} signature above is particularly simple and instructive: $u(t)=0 \Rightarrow y(t)=0$ for all $t$ and all admissible inputs. This is a behavioral constraint in the sense of Section~2, and the output form $y=g(x)\,u$ is an elementary realization, with the factor of $u$ enforcing the constraint identically
(and moreover, this property is broadly consistent with startle-habituation data from individual ciliates \cite{Rajan2023} and flies \cite{boon2026_bayesian_flies}). 
We emphasize that the analogy is methodological rather than substantive: memristive systems and habituation are associated with different behavioral constraints and thus lead to different realizations.
\end{textbox}

\subsubsection{Neuromorphic devices: elicited versus intrinsic recovery}
Across physical memristive devices, response decrement is readily achieved using diverse stimulus types; what differs is recovery: whether return toward baseline must be \emph{elicited} or occurs \emph{intrinsically} once stimulation stops. $\mathrm{SmNiO_3}$~\citep{Zuo2017} and NiO~\citep{Zhang2021resistance} exhibit habituation-like behavior under pulsatile hydrogen exposure, as do nickelates~\citep{Patel2026_NickelateEMRadiation} under UV stimulation, but recovery within experimental timescales is not automatic: it requires altering the material environment or applying a counter-stimulus (e.g., in \citep{Zhang2021resistance}, a dishabituating ozone pulse that restores the response; $H_8$) rather than simply withholding stimulation. A recent study ~\citep{park2025experimental} notes this distinction, which can be subtle in practice, and adopts the term \emph{pseudo} spontaneous recovery. Genuine intrinsic recovery, however, also occurs. Volatile memristive synapses have been used to build sensory systems that emulate several hallmarks (including spontaneous recovery), with applications to robot navigation~\citep{wu2020_material_hab_volatile}. And using the same parent material as~\citep{Zhang2021resistance} but with a different stimulus modality, Mondal et al.~\citep{mondal2022all} report all-electric habituation with resistance relaxing to baseline when stimulation is simply withheld. That intrinsic and elicited recovery appear even within a single material family underscores that the distinction is a property of device and stimulus together, not of material alone. Notably, the same physical fact --- a state relaxing toward baseline --- can read as spontaneous recovery ($H_2$) under the habituation lens, but as unwanted volatility (loss of stored information) under a memory-engineering one~\citep[Fig.~S2]{park2025experimental}. 

\subsubsection{Synthetic biology}
We return to living systems, now engineered rather than observed. 
Pla-Mauri and Sol\'e~\citep{sole2026acsSynthBiol}
argue that genetic circuits built from standard parts can exhibit not only habituation but also sensitization and massed--spaced learning, which has been observed even in non-neural cells~\citep{kukushkin2024massed}. While their habituation circuit sits between the minimal motif and models of \cite{eckert2024} in complexity, the core is shared: a repressor accumulates as a leaky molecular memory of past input, and gates the output through saturating repression. Complementing these designs, Bonzanni et al.~\citep{Bonzanni2020} show that the membrane voltage of optogenetically stimulated HEK cells reproduces the core hallmarks (decrement, recovery on rest; $H_1$, $H_2$) as well as features suggestive of frequency sensitivity $H_4$. They implicate native K\textsuperscript{+}-channel dynamics as one mediator of habituation, and demonstrate that even a non-neuronal human cell can habituate under suitably engineered control signals.
 
Together, the examples above make the notion of substrate-independence concrete: the same memory-driven attenuation can realize habituation whether the memory variable is a voltage, a material state, or a molecular concentration. In each case, the circuit is something to be \emph{built}; we turn next to a complementary role, in which the same low-dimensional structure is fit to stochastic behavioral data to infer and rationalize individual variability.

\subsection{Inferring latent dynamics from behavior}
\label{sec:inference_biology}

\begin{figure}[ht]
\includegraphics[width=0.99\linewidth]{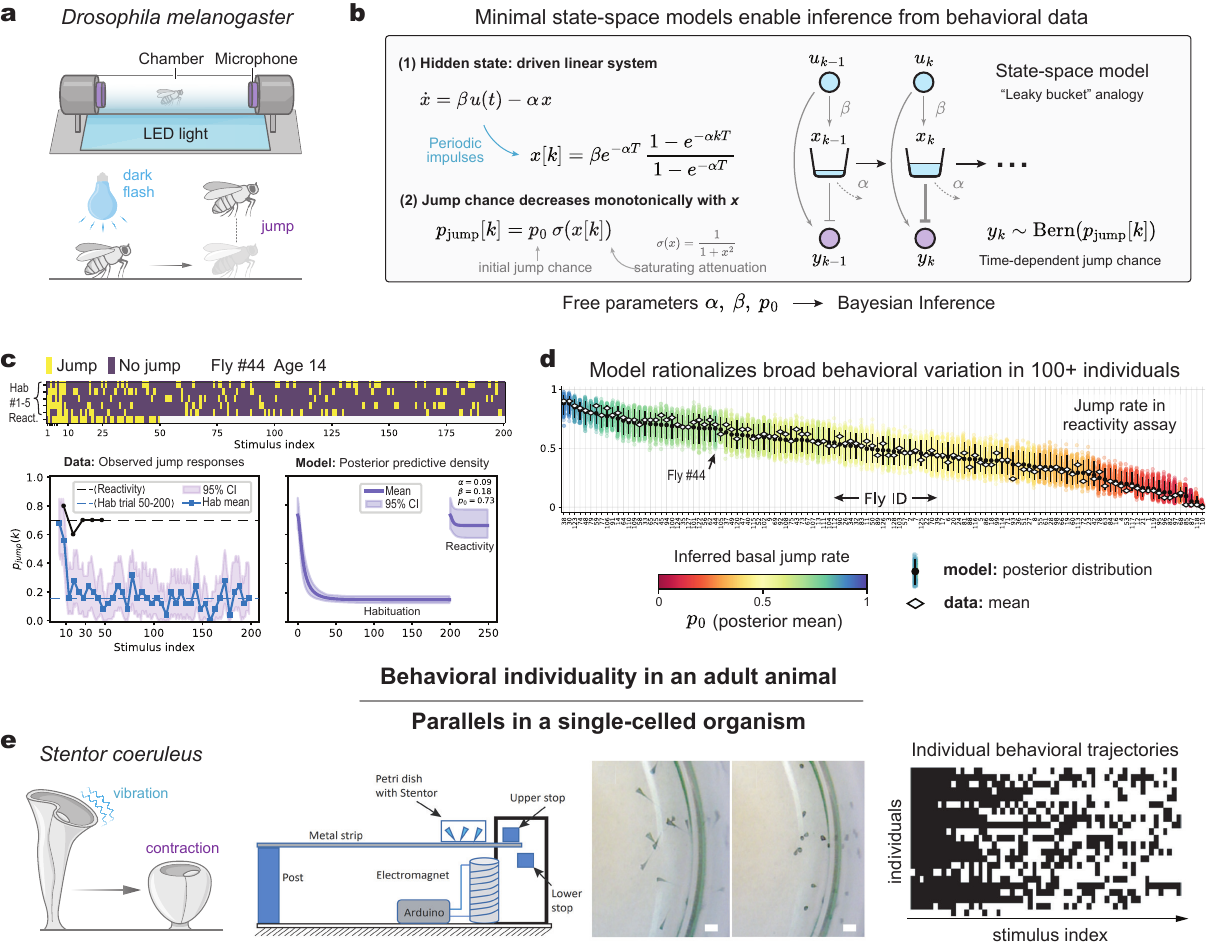}
\caption{
  \textbf{Inference of habituation dynamics across biological scales.}
  (a) Repeated stimuli drive unobserved internal dynamics that shape behavioral responses. 
  (b) A minimal state-space model represents habituation through a fading-memory variable $x$ with response probability determined by three parameters, enabling Bayesian inference from behavioral time series. 
  (c) Example fit to the responses of an individual fly, showing moving average of binary jump events and posterior predictive trajectories.
  (d) Inferred parameters reveal broad inter-individual variability across more than one hundred flies. 
  (e) Parallel measurements in the ciliate \emph{Stentor coeruleus}. Mechanical stimulation elicits contraction responses that attenuate with repetition, generating heterogeneous trajectories across individual cells. Such datasets have motivated new modeling approaches, illustrating how individual-level behavioral measurements can serve as a bridge between observed behavior and latent dynamics.
  Panels (a–d) adapted from Boon et al.\ \cite{boon2026_bayesian_flies} with permission (CC BY-NC-ND 4.0).
  Panel (e) adapted from Rajan et al.\ \cite{Rajan2023} (CC BY 4.0).}
\label{fig:fig_flyjump_SSM}
\end{figure}

Much of the experimental data underlying habituation-like behavior in biological systems reports population-averaged responses. 
This smoothing lends itself to deterministic models as discussed in the preceding sections, but obscures a basic fact: real behavior, habituation or otherwise, is stochastic within and across individuals (even when controlling for confounds like genetics and environment). As high-throughput individual-level behavioral data become increasingly available with modern experimental methods, the minimal motif acquires a second role: not as a mechanism to be built, but as a compact model class for inference. We highlight two examples that pose complementary inferential problems: a whole organism, where the model class is taken as given and one infers parameters within it, and a single cell, where the open question is which model class is right at all.

\emph{Inference within a given class: the fly.} Boon et al.~\citep{boon2026_bayesian_flies} fit a stochastic state-space extension of the minimal motif to 1,050 binary jump responses per individual fly in a light-off dark-flash assay (\textbf{Figure~\ref{fig:fig_flyjump_SSM}a-d}). Because the motif carries only three parameters, Bayesian inference is tractable at the level of single trajectories rather than population averages: each fly is summarized by inferred reactivity and habituation parameters, and the posteriors reveal substantial but structured inter-individual variation that is stable across weeks of lifespan. That a three-parameter state-space model captures much of the structure in such noisy binary data is itself notable. The approach places habituation within a broader program, exemplified by state-space and switching-dynamics models~\citep{linderman2017bayesian}, of making predictions from noisy and incomplete observations.
Of note, behavioral plasticity in the same organism has been independently modeled with low-dimensional latent dynamics by other groups~\cite{gibson2015behavioral, berne2023mechanical}; characterizing \emph{individuality} (see \cite{boon2026_bayesian_flies} and refs. therein) remains an interesting direction.
Two caveats should be noted: the approach assumes a fixed model class and infers parameters within it; and a good trajectory fit does not establish that the inferred latent variables correspond to any particular molecular mechanism. The fitted motif is, in this sense, a useful phenomenological tool rather than a mechanistic claim.


\emph{Discriminating among classes: single cells.} 
In simpler systems the prospect of a genuinely mechanistic model seems more within reach. Recent work on the giant ciliate \emph{Stentor coeruleus} is an instructive case. An elegant study by Rajan et al.~\citep{Rajan2023} tracked contraction responses of individual cells to repeated mechanical stimulation and found notably heterogeneous trajectories, which they first described (following standard practice in animal behavior) with a low-dimensional hidden Markov model involving two latent states (responsive and non-responsive). 
Subsequent work from the same group~\citep{Rajan2025_marshall} moved toward a continuous-time receptor-inactivation model of the kind discussed in Section~\ref{sec:related-models}, in which habituation emerges from the dynamics of receptor availability. 
The shift is the conceptual one this review has emphasized: an HMM infers discrete latent states with transition rates that summarize observed behavior, whereas the \emph{dynamical} model they develop in the later study posits a continuous mechanism whose parameters, once inferred, predict responses under novel stimulation schedules without re-estimation. The same move distinguishes the Boon et al.\ continuous motif from the purely descriptive population fits of earlier work.

Most recently, molecular perturbation has begun to constrain the admissible mechanisms directly~\citep{rajan2026molecular}: habituation in \emph{Stentor} appears to depend on fast calcium-channel dynamics rather than protein synthesis (consistent with its timescale) and, quite interestingly, the habituated state can persist through cell division, hinting at inheritance and community-level effects that remain largely unexplored. A full account is beyond the review's scope, but the trajectory of these studies, from HMMs to transient memory to molecular constraints, parallels the behavior-to-dynamical-mechanism angle emphasized here. Whether inference methods of the kind applied in \emph{Drosophila} \cite{boon2026_bayesian_flies} could illuminate individual-level variability in aneural systems such as \emph{Stentor} is an interesting open direction.

\subsection{Filtering, forgetting, and transient computation in machine learning}
\label{sec:ml-connections}

Natural and artificial learning systems alike must compute on a stream of input under finite resources (metabolic budget in one case, memory and compute in the other) and both must decide, on the fly, which recent input to retain. A recurring structural answer is fading memory: a state that integrates recent input and forgets the remote past. The motif developed here is the minimal instance, a one-dimensional state-space model (\textbf{Figure~\ref{fig:fig_flyjump_SSM}b}); reservoir computers and modern state-space models can be viewed as high-dimensional generalizations of related architectural principles. We trace this lineage outward, marking where the thread holds and where it gives way to looser mechanisms.

\subsubsection{Reservoir computing}
The Boyd–Chua realization of fading memory \cite{BoydChua1985fading} has a network counterpart. Maass and Sontag show that feedforward networks with dynamic synapses, even with a single hidden layer, can approximate any time-invariant fading-memory operator \citep[Thm. 1]{maass_and_sontag2000_neural_as_nonlin_filter}.
This anticipates reservoir computing, formalized in liquid state machines \citep{maass2002_liquidstatemachine} and echo state networks \citep{jaeger2001echo, jaeger2007_leakyESN}, in which a high-dimensional recurrent fading-memory network is the substrate for learning while only a static readout is trained \cite{jaeger2007special}. 
The habituating motif sits at the simplest end of this spectrum: one linear fading-memory unit with a static nonlinear readout, designed to meet behavioral constraints rather than to approximate arbitrary operators. Both rely on fading memory: for habituation, the forgetting that underlies spontaneous recovery ($H_2$); for reservoirs, the \emph{echo state} property, which Wiener earlier termed ``asymptotic independence of the remote past'' \cite{wiener1958nonlinear, BoydChua1985fading}. A notable distinction is that echo-state networks typically place the nonlinearity in the recurrence and train a linear readout \citep{jaeger2007_leakyESN}, whereas the habituation motif uses linear dynamics with nonlinear readout.
This second arrangement reappears in modern state-space models, discussed next.

\subsubsection{Selective state-space models (SSMs) and gated memory}
HiPPO \citep{gu2020hippo} introduced an online method for maintaining finite-dimensional representations of input signals through polynomial-basis projections. Building on this, Gu et al.~\citep{gu2022efficiently} introduced \emph{structured} SSMs (S4), showing that networks based on trained high-dimensional LTI layers can rival transformers on long-range benchmarks. By relaxing time-invariance, Mamba \citep{gu2024mamba} introduced \emph{selectivity}, using input-dependent state-space parameters and a timestep $\Delta(u_t)$ that modulates the recurrence's decay rate based on the current signal. We read selective SSMs not as realizations of habituation circuits but as independent instances of input-conditioned fading memory. Input-dependent gating is older still, present in LSTMs \citep{hochreiter1997long, gers2000learning} and revisited in xLSTM \citep{beck2024xlstm}. A precise connection to \emph{linear} attention mechanisms has since been established \citep{dao2024transformersAreSSMs}, and related work grounds these architectures in fading-memory and Volterra theory \citep{wang2023statespace, boyd1984analytical}; see also \citep{wang2024inverse, orvieto2023universality, cirone2024theoretical}. What recurs is not the motif but its defining move --- recent input shaping retention  --- reached independently by separate literatures.

\subsubsection{Where the thread ends: attention}
With attention-based architectures \cite{vaswani2017attention}, the thread changes character. The two mechanisms are structurally distinct. Habituation is generally understood as \emph{nonassociative} learning, while attention layers have been interpreted through the lens of \emph{associative} memory \citep{ramsauer2021hopfield, smart2025context, smart2026attentionincontextempiricalbayes}. Moreover, transformers are not a fading-memory architecture: their dependence on context does not decay with distance, by design. Fading-memory systems make the recent past available through a continuously evolving state; attention makes it available through content-addressable retrieval from the context.
Yet despite these architectural differences, resource constraints push practical deployments toward \emph{finite} context windows that reintroduce the same retention problem:
under limited memory and compute, mechanisms such as gating, pruning, or sliding windows must decide what remains available as new input arrives. 
As models scale, \emph{hybrid} architectures \citep{de2024griffin, lenz2025jamba} increasingly interleave recurrent or state-space components with attention, pairing efficient fading-memory dynamics with downstream storage and retrieval of salient information; interestingly, this integration may be more expressive than either component alone \cite{merrill2026olmohybridtheorypractice}.

Taken together, these examples suggest a recurring computational strategy: useful computation can be carried out in the \emph{transient} response of a driven dynamical system, without requiring modification of the underlying parameters or convergence to a steady state. In that sense, habituation can be viewed as an elementary member of a much broader family of transient computations.

\section{OPEN QUESTIONS}
\label{sec:open}

We close by highlighting some open questions this review raises for systems and control.  

\begin{enumerate}

  \item \textbf{Normal forms and internal models. }
  Given a system that habituates, can a universal low-dimensional habituating subsystem be isolated? This question is reminiscent of the search for bifurcation normal forms: the lowest-dimensional system implementing the desired behavior, to which all complex systems can systematically be related. This involves reducing to core variables (e.g., a Lyapunov-Schmidt reduction~\cite{golubitsky1985singularities}) and mapping these core variables to a topologically equivalent minimal description (such as a saddle-node  normal form~\cite{Kuznetsov2023}).   While recurring motifs have long been known in systems biology~\cite{Tyson2003-SniffersBuzzersEtc}, it is an open question whether such universal principles govern them. 
  Similarly, the habituation literature repeatedly invokes subtractive or predictive mechanisms (e.g., negative image models \cite{ramaswami2014network, Shen2020_odor_negative_image_model} or novelty filters \cite{textbook_kohonen_1989}) that attenuate recurring input by representing stimulus history. These resemble the Internal Model Principle for adaptation \cite{FrancisWonham1976_IMP, sontag2022_annurev_internalmodels, araujo2023universal}, raising the question of whether an analogue of the IMP for habituation exists. 
  The minimal motif and its variants do not contain an internal model of a pulse train, however. This does not contradict the IMP, since perfect rejection is not achieved (here equivalent to $\lim_{t\rightarrow\infty} y(t)= 0$). We conjecture an internal model would be necessary for perfect rejection but note that partial substantial rejection is sufficient or even advantageous in organisms.
  Other authors have argued that requiring an internal model for signals more complex than steps results in unrealistic calibration requirements for the internal model and, as a consequence, living systems are likely to adopt different mechanisms than the IMP~\cite{Sepulchre2026}.

  \item \textbf{Behavioral identification and model discovery.} 
  Given observations of a habituating system, how can one infer not only parameters but also the underlying dynamical structure? Existing approaches occupy opposite ends of the spectrum: system identification, symbolic regression, and common Bayesian methods fit models to measured trajectories, while the present review derives models from qualitative behavioral constraints. Discriminating amongst distinct model classes using both data and constraints remains a challenge; the behavioral framework of Willems \citep{PoldermanWillems1998,willems2007behavioral} provides a natural language here, but whether it can be extended to support qualitative constraints on transient response appears open.
  Another twist is that the hallmarks are formulated deterministically, yet biological realizations are noisy. What does it mean for a transient behavioral constraint to hold probabilistically, and can robust guarantees analogous to those known for perfect adaptation \citep{Briat201615,Khammash2019} be established?

  \item \textbf{Compositionality and MIMO behavior.} 
  We have focused on a single SISO unit, but its extensions are already compositional: series connection yields frequency-dependent recovery $H_{4\textrm{(b)}}$ \cite{Staddon1993, Staddon1996}, and parallel arrangements have been used for spatial novelty detection \cite{textbook_kohonen_1989}. Is there a path to composing behavioral constraints, analogous to the interconnection rules for input–output systems in the behavioral framework \cite{willems2007behavioral}? 
  Furthermore, which hallmarks are preserved under series, parallel, or feedback composition, and which require a dedicated structure? 
  This is also where the multivariable hallmarks re-enter: in the MIMO setting, novelty filter and negative-image perspectives \cite{textbook_kohonen_1989, ramaswami2014network, Shen2020_odor_negative_image_model, bourassa2026prxlife} can encode relationships among inputs, potentially blurring the line between nonassociative and associative learning.

\end{enumerate}

\section{CONCLUSION}
\label{sec:conclusions}

This review framed habituation as a problem of designing a dynamical system: given qualitative constraints on a system's transient response, identify the minimal dynamical ingredients that meet them. The general framing yields firm results: nonnegative, bounded habituation rules out linearity, making nonlinearity structural rather than optional, and a single fading-memory state with a static nonlinear readout already captures the core hallmarks. Together with simple extensions of this core motif, the perspective organizes an otherwise scattered literature by structure rather than biological origin.

The framing also reconnects, at the end, to the filtering role with which it begins: as a cheap front-end gate that attenuates the familiar, habituation controls what reaches slower, more elaborate downstream processes. Sections 2-5 largely concern the transient gating itself (its structure, its hallmarks, its realizations), and the resulting synthesis has little bearing on what lies downstream. In particular, habituation is one concrete instance of transient information storage in dynamical systems \cite{ganguli2008memory}, but how such storage interfaces with longer-lived memory remains poorly understood \cite{ramaswami2014network}. The ubiquity of habituation is often attributed to its role as a sensory firewall that guards finite cognitive and metabolic resources \cite{poon2006nonassociative}, and possibly shapes what becomes available for consolidation and learning \cite{dudai2015consolidation}, though whether the filter merely shields later machinery or actively selects for it is unresolved. If habituation is in part a resource-allocation mechanism regulating access to downstream circuitry, the design question generalizes: might analogous filtering principles prove useful in artificial learning systems under their own resource constraints? That so simple a filter sits at the entrance to such different downstream machinery suggests the gating it performs is a shared design question rather than a biological idiosyncrasy --- but establishing that, and characterizing the handoff from fading attenuation to durable storage across biological, physical, and computational systems, remains an open challenge.

\section*{DISCLOSURE STATEMENT}
The authors are not aware of any affiliations, memberships, funding, or financial holdings that might be perceived as affecting the objectivity of this review.

\section*{ACKNOWLEDGMENTS}
The authors thank Jeremy Gunawardena for drawing their attention to relevant questions and literature, and Annette Schenck, Bob Carpenter, Fran\c{c}ois Bourassa, Nikolay Kukushkin, Shriram Ramanathan, and Wallace Marshall for valuable discussions and suggestions. They are particularly grateful to Eduardo D. Sontag for feedback of various kinds. They also thank Rodolphe Sepulchre and an anonymous reviewer for comments that improved the manuscript. M.M. acknowledges travel support from the Simons Foundation.

\textbf{\\Significance statement (120 words)}
Living systems must distinguish meaningful signals from background distractions to survive and reproduce. Habituation, the progressive loss of response to repeated, harmless stimuli and its recovery once they stop, is the simplest example of this filtering. It is observed across animals and many other living systems, and even certain nonliving materials. Despite its universality, the dynamical ingredients required for habituation remain incompletely formalized. We ask: given qualitative constraints imposed by habituation on a system's response, what is the minimal dynamical structure that satisfies them? Drawing on systems and control theory, we identify a small set of structural motifs that capture the classical hallmarks of habituation and connect them to work in biology, engineered devices, and machine learning.



\begin{thebibliography}{130}
\expandafter\ifx\csname natexlab\endcsname\relax\def\natexlab#1{#1}\fi

\bibitem{smart2024pnas}
Smart M, Shvartsman SY, Mönnigmann M. 2024.
Minimal motifs for habituating systems.
\textit{Proceedings of the National Academy of Sciences} 121(41):e2409330121

\bibitem{Monnigmann2024CDC}
Smart M, Shvartsman SY, M\"onnigmann M. 2024.
\textit{A minimal dynamical system and analog circuit for non-associative learning}.
In \textit{2024 IEEE 63rd Conference on Decision and Control (CDC)}, pp.  577--582. Milan, Italy: IEEE

\bibitem{Thompson2009}
Thompson RF. 2009.
Habituation: A history.
\textit{Neurobiology of learning and memory} 92(2):127--134

\bibitem{jennings1902studies}
Jennings HS. 1902.
Studies on reactions to stimuli in unicellular organisms. ix.—on the behavior of fixed infusoria ({Stentor} and {Vorticella}), with special reference to the modifiability of protozoan reactions.
\textit{American Journal of Physiology} 8(1):23--60

\bibitem{Jennings1906}
Jennings HS. 1906.
\textit{Behavior of the lower organisms}.
Columbia University Press

\bibitem{davis1970effects}
Davis M. 1970.
Effects of interstimulus interval length and variability on startle-response habituation in the rat.
\textit{Journal of Comparative and Physiological Psychology} 72(2):177--192

\bibitem{groves1970habituation}
Groves PM, Thompson RF. 1970.
Habituation: a dual-process theory.
\textit{Psychological Review} 77(5):419--450

\bibitem{Randlett2019}
Randlett O, Haesemeyer M, Forkin G, Shoenhard H, Schier AF, et~al. 2019.
Distributed plasticity drives visual habituation learning in larval zebrafish.
\textit{Current Biology} 29(8):1337--1345

\bibitem{Carew1972}
Carew TJ, Pinsker HM, Kandel ER. 1972.
Long-term habituation of a defensive withdrawal reflex in {Aplysia}.
\textit{Science} 175(4020):451--454

\bibitem{kandel1976cellular}
Kandel ER. 1976.
\textit{Cellular Basis of Behavior: An Introduction to Behavioral Neurobiology}.
W. H. Freeman

\bibitem{rankin1990caenorhabditis}
Rankin CH, Beck CD, Chiba CM. 1990.
{C}aenorhabditis elegans: a new model system for the study of learning and memory.
\textit{Behavioural brain research} 37(1):89--92

\bibitem{Engel2009}
Engel JE, Wu CF. 2009.
Neurogenetic approaches to habituation and dishabituation in {Drosophila}.
\textit{Neurobiology of Learning and Memory} 92(2):166--175

\bibitem{Mcfadden1990}
McFadden PN, Koshland DE. 1990.
Habituation in the single cell: Diminished secretion of norepinephrine with repetitive depolarization of {PC12} cells.
\textit{Proceedings of the National Academy of Sciences of the United States of America} 87(5):2031--2035

\bibitem{Cheever1994}
Cheever L, Koshland DE. 1994.
Habituation of neurosecretory responses to extracellular {ATP} in {PC12} cells.
\textit{Journal of Neuroscience} 14(8):4831--4838

\bibitem{Gagliano2014}
Gagliano M, Renton M, Depczynski M, Mancuso S. 2014.
Experience teaches plants to learn faster and forget slower in environments where it matters.
\textit{Oecologia} 175(1):63--72

\bibitem{ortega1970phycomyces}
Ortega JK, Gamow RI. 1970.
Phycomyces: habituation of the light growth response.
\textit{Science} 168(3937):1374--1375

\bibitem{applewhite1975learning}
Applewhite PB. 1975.
Learning in bacteria, fungi, and plants.
\textit{Invertebrate Learning} 3:179--186

\bibitem{wood1969parametric}
Wood DC. 1969.
Parametric studies of the response decrement produced by mechanical stimuli in the protozoan, {Stentor} coeruleus.
\textit{Journal of neurobiology} 1(3):345--360

\bibitem{Eisenstein1982}
Eisenstein EM, Brunder DG, Blair HJ. 1982.
Habituation and sensitization in an aneural cell: Some comparative and theoretical considerations.
\textit{Neuroscience and Biobehavioral Reviews} 6(2):183--194

\bibitem{Rajan2023}
Rajan D, Makushok T, Kalish A, Acuna L, Bonville A, et~al. 2023.
Single-cell analysis of habituation in {Stentor} coeruleus.
\textit{Current Biology} 33(2):241--251

\bibitem{Boisseau2016}
Boisseau RP, Vogel D, Dussutour A. 2016.
Habituation in non-neural organisms: Evidence from slime moulds.
\textit{Proceedings of the Royal Society B: Biological Sciences} 283(1829):20160446

\bibitem{Boussard2019}
Boussard A, Delescluse J, Pérez-Escudero A, Dussutour A. 2019.
Memory inception and preservation in slime moulds: The quest for a common mechanism.
\textit{Philosophical Transactions of the Royal Society B: Biological Sciences} 374(1774):20180368

\bibitem{Zhang2021resistance}
Zhang Z, Mondal S, Mandal S, Allred JM, Aghamiri NA, et~al. 2021.
Neuromorphic learning with {Mott} insulator {NiO}.
\textit{Proceedings of the National Academy of Sciences} 118(39):e2017239118

\bibitem{wu2020_material_hab_volatile}
Wu Z, Lu J, Shi T, Zhao X, Zhang X, et~al. 2020.
A habituation sensory nervous system with memristors.
\textit{Advanced Materials} 32(46):2004398

\bibitem{gershman2021reconsidering}
Gershman SJ, Balbi PE, Gallistel CR, Gunawardena J. 2021.
Reconsidering the evidence for learning in single cells.
\textit{eLife} 10:e61907

\bibitem{gunawardena2022learning}
Gunawardena J. 2022.
Learning outside the brain: Integrating cognitive science and systems biology.
\textit{Proceedings of the IEEE} 110(5):590--612

\bibitem{Thompson1966}
Thompson RF, Spencer WA. 1966.
Habituation: a model phenomenon for the study of neuronal substrates of behavior.
\textit{Psychological Review} 73(1):16--43

\bibitem{rankin2009}
Rankin CH, Abrams T, Barry RJ, Bhatnagar S, Clayton DF, et~al. 2009.
{Habituation revisited: An updated and revised description of the behavioral characteristics of habituation}.
\textit{Neurobiology of Learning and Memory} 92(2):135--138

\bibitem{eckert2024}
Eckert L, Vidal-Saez MS, Zhao Z, Garcia-Ojalvo J, Martinez-Corral R, Gunawardena J. 2024.
Biochemically plausible models of habituation for single-cell learning.
\textit{Current Biology} 34(24):5646--5658.e3

\bibitem{Wiener1948cybernetics}
Wiener N. 1948.
\textit{Cybernetics: Or Control and Communication in the Animal and the Machine}.
Cambridge, MA: The Technology Press / Wiley

\bibitem{ashby1956introductionCybernetics}
Ashby WR. 1956.
\textit{An Introduction to Cybernetics}.
London: Chapman \& Hall

\bibitem{Blok2022}
Blok LER, Boon M, van Reijmersdal B, Höffler KD, Fenckova M, Schenck A. 2022.
Genetics, molecular control and clinical relevance of habituation learning.
\textit{Neuroscience and Biobehavioral Reviews} 143:104883

\bibitem{Fenckova2019}
Fenckova M, Blok LE, Asztalos L, Goodman DP, Cizek P, et~al. 2019.
Habituation learning is a widely affected mechanism in {D}rosophila models of intellectual disability and autism spectrum disorders.
\textit{Biological psychiatry} 86(4):294--305

\bibitem{poon2006nonassociative}
Poon CS, Young DL. 2006.
Nonassociative learning as gated neural integrator and differentiator in stimulus-response pathways.
\textit{Behavioral and Brain Functions} 2(1):29

\bibitem{Ljung1999-SysIdTheoryForTheUser}
Ljung L. 1999.
\textit{System Identification: Theory for the User}.
Prentice Hall information and system sciences series. Prentice Hall PTR

\bibitem{Ljung2008-PerspectivesOnSystemsIdentification}
Ljung L. 2010.
Perspectives on system identification.
\textit{Annual Reviews in Control} 34(1):1--12

\bibitem{Ferrell201662}
Ferrell JE. 2016.
Perfect and near-perfect adaptation in cell signaling.
\textit{Cell Systems} 2(2):62--67

\bibitem{Tyson2003-SniffersBuzzersEtc}
Tyson JJ, Chen KC, Novak B. 2003.
Sniffers, buzzers, toggles and blinkers: dynamics of regulatory and signaling pathways in the cell.
\textit{Current Opinion in Cell Biology} 15(2):221--231

\bibitem{Briat201615}
Briat C, Gupta A, Khammash M. 2016.
Antithetic integral feedback ensures robust perfect adaptation in noisy biomolecular networks.
\textit{Cell Systems} 2(1):15--26

\bibitem{brunton2016discovering}
Brunton SL, Proctor JL, Kutz JN. 2016.
Discovering governing equations from data by sparse identification of nonlinear dynamical systems.
\textit{Proceedings of the National Academy of Sciences} 113(15):3932--3937

\bibitem{schmidt2009distilling}
Schmidt M, Lipson H. 2009.
Distilling free-form natural laws from experimental data.
\textit{Science} 324(5923):81--85

\bibitem{GuckenheimerHolmes2002}
Guckenheimer J, Holmes P. 2002.
\textit{Nonlinear Oscillations, Dynamical Systems, and Bifurcations of Vector Fields}, vol.~42 of \textit{Applied Mathematical Sciences}.
Springer-Verlag, 7th ed.

\bibitem{golubitsky1985singularities}
Golubitsky M, Schaeffer DG. 1985.
\textit{Singularities and Groups in Bifurcation Theory: Volume I}, vol.~51 of \textit{Applied Mathematical Sciences}.
New York, NY: Springer-Verlag

\bibitem{PoldermanWillems1998}
Polderman JW, Willems JC. 1998.
\textit{Introduction to mathematical systems theory}.
Texts in applied mathematics: 26. Springer New York, NY

\bibitem{willems2007behavioral}
Willems JC. 2007.
The behavioral approach to open and interconnected systems.
\textit{IEEE control systems magazine} 27(6):46--99

\bibitem{sepulchre2018excitable}
Sepulchre R, Drion G, Franci A. 2018.
Excitable behaviors.
In \textit{Emerging Applications of Control and Systems Theory}. Springer

\bibitem{ribar2021neuromorphic}
Ribar L, Sepulchre R. 2021.
Neuromorphic control: Designing multiscale mixed-feedback systems.
\textit{IEEE Control Systems Magazine} 41(6):34--63

\bibitem{Ma2009}
Ma W, Trusina A, El-Samad H, Lim WA, Tang C. 2009.
Defining network topologies that can achieve biochemical adaptation.
\textit{Cell} 138(4):760--773

\bibitem{yi2000robust}
Yi TM, Huang Y, Simon MI, Doyle J. 2000.
Robust perfect adaptation in bacterial chemotaxis through integral feedback control.
\textit{Proceedings of the National Academy of Sciences} 97(9):4649--4653

\bibitem{tu2013quantitative}
Tu Y. 2013.
Quantitative modeling of bacterial chemotaxis: signal amplification and accurate adaptation.
\textit{Annual Review of Biophysics} 42:337--359

\bibitem{tu2018adaptation}
Tu Y, Rappel WJ. 2018.
Adaptation in living systems.
\textit{Annual Review of Condensed Matter Physics} 9:183--205

\bibitem{Khammash2019}
Aoki SK, Lillacci G, Gupta A, Baumschlager A, Schweingruber D, Khammash M. 2019.
{A universal biomolecular integral feedback controller for robust perfect adaptation}.
\textit{Nature} 570(7762):533--537

\bibitem{FrancisWonham1976_IMP}
Francis BA, Wonham WM. 1976.
The internal model principle of control theory.
\textit{Automatica} 12(5):457--465

\bibitem{sontag2022_annurev_internalmodels}
Bin M, Huang J, Isidori A, Marconi L, Mischiati M, Sontag E. 2022.
Internal models in control, bioengineering, and neuroscience.
\textit{Annual Review of Control, Robotics, and Autonomous Systems} 5(1):55--79

\bibitem{Farina2000}
Farina L, Rinaldi S. 2000.
\textit{Positive {L}inear {S}ystems: {T}heory and {A}pplications}.
Pure and applied mathematics. New York: John Wiley \& Sons

\bibitem{boyd1984analytical}
Boyd S, Chua LO, Desoer CA. 1984.
Analytical foundations of {V}olterra series.
\textit{IMA Journal of Mathematical Control and Information} 1(3):243--282

\bibitem{maass_and_sontag2000_neural_as_nonlin_filter}
Maass W, Sontag ED. 2000.
Neural systems as nonlinear filters.
\textit{Neural Computation} 12(8):1743--1772

\bibitem{Schoukens2017}
Schoukens M, Tiels K. 2017.
Identification of block-oriented nonlinear systems starting from linear approximations: A survey.
\textit{Automatica} 85:272--292

\bibitem{ramaswami2014network}
Ramaswami M. 2014.
Network plasticity in adaptive filtering and behavioral habituation.
\textit{Neuron} 82(6):1216--1229

\bibitem{Shen2020_odor_negative_image_model}
Shen Y, Dasgupta S, Navlakha S. 2020.
Habituation as a neural algorithm for online odor discrimination.
\textit{Proceedings of the National Academy of Sciences} 117(22):12402--12410

\bibitem{Gunawardena2005}
Gunawardena J. 2005.
Multisite protein phosphorylation makes a good threshold but can be a poor switch.
\textit{Proceedings of the National Academy of Sciences} 102(41):14617--14622

\bibitem{Staddon1993}
Staddon JE. 1993.
On rate-sensitive habituation.
\textit{Adaptive Behavior} 1(4):421--436

\bibitem{Staddon1996}
Staddon JE, Higa JJ. 1996.
Multiple time scales in simple habituation.
\textit{Psychological Review} 103(4):720--733

\bibitem{staddon2001adaptive}
Staddon JE. 2001.
\textit{Adaptive Dynamics: The Theoretical Analysis of Behavior}.
Cambridge, MA: The MIT Press

\bibitem{Stanley1976}
Stanley JC. 1976.
Computer simulation of a model of habituation.
\textit{Nature} 261(5556):146--148

\bibitem{DLWang1993}
Wang D. 1993.
A neural model of synaptic plasticity underlying short-term and long-term habituation.
\textit{Adaptive Behavior} 2(2):111--129

\bibitem{Dragoi2002}
Dragoi V. 2002.
A feedforward model of suppressive and facilitatory habituation effects.
\textit{Biological cybernetics} 86(6):419--426

\bibitem{delrosal2006}
{del Rosal} E, Alonso L, Moreno R, Vázquez M, Santacreu J. 2006.
Simulation of habituation to simple and multiple stimuli.
\textit{Behavioural Processes} 73(3):272--277

\bibitem{tsodyks1997neural}
Tsodyks MV, Markram H. 1997.
The neural code between neocortical pyramidal neurons depends on neurotransmitter release probability.
\textit{Proceedings of the National Academy of Sciences} 94(2):719--723

\bibitem{Aljadeff2016}
Aljadeff J, Lansdell BJ, Fairhall AL, Kleinfeld D. 2016.
Analysis of neuronal spike trains, deconstructed.
\textit{Neuron} 91(2):221--259

\bibitem{WeberFairhall2019}
Weber AI, Fairhall AL. 2019.
The role of adaptation in neural coding.
\textit{Current Opinion in Neurobiology} 58:135--140

\bibitem{Sokolov1963perception}
Sokolov E. 1963.
\textit{Perception and the Conditioned Reflex}.
Pergamon Press book. Pergamon Press

\bibitem{textbook_kohonen_1989}
Kohonen T. 1989.
\textit{Self-organization and associative memory: 3rd edition}.
Berlin, Heidelberg: Springer-Verlag

\bibitem{bourassa2026prxlife}
Bourassa FXP, Fran{\c{c}}ois P, Reddy G, Vergassola M. 2026.
Manifold learning for olfactory habituation to strongly fluctuating backgrounds.
\textit{PRX Life} 4(1):013008

\bibitem{sole2026acsSynthBiol}
Pla-Mauri J, Sol{\'e} R. 2026.
Engineering basal cognition: Minimal genetic circuits for habituation, sensitization, and massed--spaced learning.
\textit{ACS Synthetic Biology} 15(2):716--727

\bibitem{Bonzanni2019}
Bonzanni M, Rouleau N, Levin M, Kaplan DL. 2019.
On the generalization of habituation: how discrete biological systems respond to repetitive stimuli: a novel model of habituation that is independent of any biological system.
\textit{Bioessays} 41(7):1900028

\bibitem{nicoletti-busiello-2024}
Nicoletti G, Bruzzone M, Suweis S, Dal~Maschio M, Busiello DM. 2025.
Optimal information gain at the onset of habituation to repeated stimuli.
\textit{eLife} 13:RP99767

\bibitem{gershman_2024}
Gershman SJ. 2024.
Habituation as optimal filtering.
\textit{iScience} 27(8):110523

\bibitem{Komatsu2025}
Komatsu M, Yasui T, Ohkawa T, Budd C. 2025.
Data-driven modeling of habituation with its frequency-dependent hallmark based on {F}ourier neural operator.
\textit{Nonlinear Theory and Its Applications, IEICE} 16(3):461--479

\bibitem{BoydChua1985fading}
Boyd S, Chua LO. 1985.
Fading memory and the problem of approximating nonlinear operators with {Volterra} series.
\textit{IEEE Transactions on Circuits and Systems} CAS-32(11):1150--1161

\bibitem{koch2024biological}
Koch D, Nandan A, Ramesan G, Koseska A. 2024.
Biological computations: limitations of attractor-based formalisms and the need for transients.
\textit{Biochemical and Biophysical Research Communications} 720:150069

\bibitem{riviere2023plants}
Rivi{\`e}re M, Meroz Y. 2023.
Plants sum and subtract stimuli over different timescales.
\textit{Proceedings of the National Academy of Sciences} 120(42):e2306655120

\bibitem{schetzen1980volterra}
Schetzen M. 1980.
\textit{The {V}olterra and {W}iener Theories of Nonlinear Systems}.
New York: Wiley

\bibitem{rugh1981nonlinear}
Rugh WJ. 1981.
\textit{Nonlinear system theory}.
Johns Hopkins University Press, Baltimore

\bibitem{bainier2026state}
Bainier G, Chaillet A, Sepulchre R, Franci A. 2026.
State-space fading memory.
\textit{arXiv preprint arXiv:2603.23814}

\bibitem{rieke1996spikes}
Rieke F, Warland D, Van~Steveninck RdR, Bialek W. 1996.
\textit{Spikes: exploring the neural code}.
MIT press

\bibitem{volterra1959theory}
Volterra V. 1959.
\textit{Theory of Functionals and of Integral and Integro-Differential Equations}.
Dover Publications

\bibitem{wiener1958nonlinear}
Wiener N. 1958.
\textit{Nonlinear Problems in Random Theory}.
Cambridge, MA: MIT Press

\bibitem{Chua1976-memristiveSystems}
Chua L, Kang SM. 1976.
Memristive devices and systems.
\textit{Proceedings of the IEEE} 64(2):209--223

\bibitem{chua1971memristor}
Chua L. 1971.
Memristor-the missing circuit element.
\textit{IEEE Transactions on Circuit Theory} 18(5):507--519

\bibitem{Pershin2009_memristive_amoeba}
Pershin YV, La~Fontaine S, Di~Ventra M. 2009.
Memristive model of amoeba learning.
\textit{Phys. Rev. E} 80(2):021926

\bibitem{boon2026_bayesian_flies}
Boon M, Smart M, Persikov AV, van Reijmersdal B, Maghbouli M, et~al. 2026.
Dynamical modeling of individual sensory reactivity and habituation learning.
\textit{Proceedings of the National Academy of Sciences} 123(13):e2524738123

\bibitem{Zuo2017}
Zuo F, Panda P, Kotiuga M, Li J, Kang M, et~al. 2017.
Habituation based synaptic plasticity and organismic learning in a quantum perovskite.
\textit{Nature Communications} 8(1):240

\bibitem{Patel2026_NickelateEMRadiation}
Patel RK, Zama K, Smart M, Eathirajan R, Seskar I, et~al. 2026.
Electromagnetic radiation stimulated learning in perovskite nickelates.
\textit{Advanced Science} :e75984

\bibitem{park2025experimental}
Park SO, Jeong H, Seo S, Kwon Y, Lee J, Choi S. 2025.
Experimental demonstration of third-order memristor-based artificial sensory nervous system for neuro-inspired robotics.
\textit{Nature Communications} 16(1):5754

\bibitem{mondal2022all}
Mondal S, Zhang Z, Islam AN, Andrawis R, Gamage S, et~al. 2022.
All-electric nonassociative learning in nickel oxide.
\textit{Advanced Intelligent Systems} 4(10):2200069

\bibitem{kukushkin2024massed}
Kukushkin NV, Carney RE, Tabassum T, Carew TJ. 2024.
The massed-spaced learning effect in non-neural human cells.
\textit{Nature Communications} 15(1):9635

\bibitem{Bonzanni2020}
Bonzanni M, Rouleau N, Levin M, Kaplan DL. 2020.
Optogenetically induced cellular habituation in non-neuronal cells.
\textit{PLoS One} 15(1):e0227230

\bibitem{linderman2017bayesian}
Linderman S, Johnson M, Miller A, Adams R, Blei D, Paninski L. 2017.
\textit{Bayesian learning and inference in recurrent switching linear dynamical systems}.
In \textit{Artificial intelligence and statistics}, pp.  914--922. PMLR

\bibitem{gibson2015behavioral}
Gibson WT, Gonzalez CR, Fernandez C, Ramasamy L, Tabachnik T, et~al. 2015.
Behavioral responses to a repetitive visual threat stimulus express a persistent state of defensive arousal in {Drosophila}.
\textit{Current Biology} 25(11):1401--1415

\bibitem{berne2023mechanical}
Berne A, Zhang T, Shomar J, Ferrer AJ, Valdes A, et~al. 2023.
Mechanical vibration patterns elicit behavioral transitions and habituation in crawling {Drosophila} larvae.
\textit{eLife} 12:e69205

\bibitem{Rajan2025_marshall}
Rajan DH, Marshall WF. 2025.
A receptor-inactivation model for single-celled habituation in {Stentor} coeruleus.
\textit{Current Biology} 35(14):3327--3340

\bibitem{rajan2026molecular}
Rajan DH, Albright A, Kim H, Diaz U, Hudnall Y, et~al. 2026.
Molecular pathways for learning in the single-cell {Stentor} coeruleus.
\textit{Current Biology} 36(9):2367--2381

\bibitem{maass2002_liquidstatemachine}
Maass W, Natschläger T, Markram H. 2002.
Real-time computing without stable states: A new framework for neural computation based on perturbations.
\textit{Neural Computation} 14(11):2531--2560

\bibitem{jaeger2001echo}
Jaeger H. 2001.
The “echo state” approach to analysing and training recurrent neural networks-with an erratum note.
\textit{German national research center for information technology GMD technical report} 148(34):13

\bibitem{jaeger2007_leakyESN}
Jaeger H, Lukoševičius M, Popovici D, Siewert U. 2007.
Optimization and applications of echo state networks with leaky-integrator neurons.
\textit{Neural Networks} 20(3):335--352

\bibitem{jaeger2007special}
Jaeger H, Maass W, Principe J. 2007.
Special issue on echo state networks and liquid state machines.
\textit{Neural Networks} 20(3):287--289

\bibitem{gu2020hippo}
Gu A, Dao T, Ermon S, Rudra A, R\'{e} C. 2020.
\textit{{HiPPO}: Recurrent Memory with Optimal Polynomial Projections}.
In \textit{Advances in Neural Information Processing Systems}, ed. H~Larochelle, M~Ranzato, R~Hadsell, M~Balcan, H~Lin, pp.  1474--1487, vol.~33, pp.  1474--1487. Curran Associates, Inc.

\bibitem{gu2022efficiently}
Gu A, Goel K, Re C. 2022.
\textit{Efficiently Modeling Long Sequences with Structured State Spaces}.
In \textit{International Conference on Learning Representations}. OpenReview.net

\bibitem{gu2024mamba}
Gu A, Dao T. 2024.
\textit{{Mamba}: Linear-Time Sequence Modeling with Selective State Spaces}.
In \textit{First Conference on Language Modeling}

\bibitem{hochreiter1997long}
Hochreiter S, Schmidhuber J. 1997.
Long short-term memory.
\textit{Neural Computation} 9(8):1735--1780

\bibitem{gers2000learning}
Gers FA, Schmidhuber J, Cummins F. 2000.
Learning to forget: Continual prediction with {LSTM}.
\textit{Neural Computation} 12(10):2451--2471

\bibitem{beck2024xlstm}
Beck M, P\"{o}ppel K, Spanring M, Auer A, Prudnikova O, et~al. 2024.
\textit{{xLSTM}: Extended Long Short-Term Memory}.
In \textit{Advances in Neural Information Processing Systems}, ed. A~Globerson, L~Mackey, D~Belgrave, A~Fan, U~Paquet, J~Tomczak, C~Zhang, pp.  107547--107603, vol.~37, pp.  107547--107603. Curran Associates, Inc.

\bibitem{dao2024transformersAreSSMs}
Dao T, Gu A. 2024.
\textit{Transformers are {SSM}s: Generalized Models and Efficient Algorithms Through Structured State Space Duality}.
In \textit{Forty-first International Conference on Machine Learning}. PMLR

\bibitem{wang2023statespace}
Wang S, Xue B. 2023.
\textit{State-space models with layer-wise nonlinearity are universal approximators with exponential decaying memory}.
In \textit{Thirty-seventh Conference on Neural Information Processing Systems}. Curran Associates, Inc.

\bibitem{wang2024inverse}
Wang S, Li Z, Li Q. 2024.
\textit{Inverse Approximation Theory for Nonlinear Recurrent Neural Networks}.
In \textit{The Twelfth International Conference on Learning Representations}

\bibitem{orvieto2023universality}
Orvieto A, De S, Gulcehre C, Pascanu R, Smith SL. 2024.
\textit{Universality of linear recurrences followed by non-linear projections: finite-width guarantees and benefits of complex eigenvalues}.
In \textit{Proceedings of the 41st International Conference on Machine Learning}, pp.  38837--38863. PMLR

\bibitem{cirone2024theoretical}
Cirone NM, Orvieto A, Walker B, Salvi C, Lyons T. 2024.
\textit{Theoretical Foundations of Deep Selective State-Space Models}.
In \textit{The Thirty-eighth Annual Conference on Neural Information Processing Systems}. Curran Associates, Inc.

\bibitem{vaswani2017attention}
Vaswani A, Shazeer N, Parmar N, Uszkoreit J, Jones L, et~al. 2017.
\textit{Attention is All you Need}.
In \textit{Advances in Neural Information Processing Systems}, ed. I~Guyon, UV~Luxburg, S~Bengio, H~Wallach, R~Fergus, S~Vishwanathan, R~Garnett, vol.~30. Curran Associates, Inc.

\bibitem{ramsauer2021hopfield}
Ramsauer H, Sch{\"a}fl B, Lehner J, Seidl P, Widrich M, et~al. 2021.
\textit{Hopfield Networks is All You Need}.
In \textit{International Conference on Learning Representations}. OpenReview.net

\bibitem{smart2025context}
Smart M, Bietti A, Sengupta AM. 2025.
\textit{In-Context Denoising with One-Layer Transformers: Connections between Attention and Associative Memory Retrieval}.
In \textit{International Conference on Machine Learning}, pp.  55950--55971. PMLR

\bibitem{smart2026attentionincontextempiricalbayes}
Smart M, Ganguly S, Metya N, Morozov AV, Sengupta AM. 2026.
Attention as in-context empirical bayes: A two-stage view via particle dynamics.
\textit{arXiv preprint arXiv:2605.29351}

\bibitem{de2024griffin}
De S, Smith SL, Fernando A, Botev A, Cristian-Muraru G, et~al. 2024.
Griffin: Mixing gated linear recurrences with local attention for efficient language models.
\textit{arXiv preprint arXiv:2402.19427}

\bibitem{lenz2025jamba}
Lenz B, Lieber O, Arazi A, Bergman A, Manevich A, et~al. 2025.
\textit{{Jamba}: Hybrid Transformer-{M}amba Language Models}.
In \textit{The Thirteenth International Conference on Learning Representations}. OpenReview.net

\bibitem{merrill2026olmohybridtheorypractice}
Merrill W, Li Y, Romero T, Svete A, Costello C, et~al. 2026.
Olmo hybrid: From theory to practice and back.
\textit{arXiv preprint arXiv:2604.03444}

\bibitem{Kuznetsov2023}
Kuznetsov Y. 2023.
\textit{Elements of Applied Bifurcation Theory}.
Applied Mathematical Sciences. Germany: Springer, 4th ed.

\bibitem{araujo2023universal}
Araujo RP, Liotta LA. 2023.
Universal structures for adaptation in biochemical reaction networks.
\textit{Nature Communications} 14(1):2251

\bibitem{Sepulchre2026}
Sepulchre R, Cecconi A, Bin M, Marconi L. 2026.
Regulation without calibration: From trajectory to event regulation.
\textit{IEEE Control Systems} 46(1):55--69

\bibitem{ganguli2008memory}
Ganguli S, Huh D, Sompolinsky H. 2008.
Memory traces in dynamical systems.
\textit{Proceedings of the National Academy of Sciences} 105(48):18970--18975

\bibitem{dudai2015consolidation}
Dudai Y, Karni A, Born J. 2015.
The consolidation and transformation of memory.
\textit{Neuron} 88(1):20--32

\end{thebibliography}

\end{document}